\documentclass[%
preprint,
superscriptaddress,
amsmath, 
amssymb,
aps,
floatfix,
]{revtex4-1}
\usepackage{lipsum, babel}
\usepackage{printlen}
\usepackage{float}
\usepackage[toc,page]{appendix}
\usepackage{graphicx}% Include figure files
\usepackage{dcolumn}% Align table columns on decimal point
\usepackage{bm}% bold math
\usepackage[colorlinks=true,citecolor=blue,linkcolor=magenta]{hyperref}
\usepackage[utf8]{inputenc}
\usepackage{url}
\usepackage[normalem]{ulem}
\usepackage{upgreek}
\usepackage[mathlines]{lineno}%Enable numbering of text and display math
\usepackage{amsmath}
\usepackage{xr}
\usepackage{verbatim}
\newcommand{\cmt}[1]{{}}
\newif\ifproofread

\usepackage[usenames,dvipsnames]{color}

\usepackage{xr}
\makeatletter
\newcommand*{\addFileDependency}[1]{% argument=file name and extension
  \typeout{(#1)}
  \@addtofilelist{#1}
  \IfFileExists{#1}{}{\typeout{No file #1.}}
}
\makeatother

\begin{document}
%\bstctlcite{IEEEexample:BSTcontrol}
\proofreadtrue

\title{Degeneracy-Locked Optical Parametric Oscillator}
%\title{Synchronized counter-propagating optical parametric oscillator}

\author{Fengyan Yang}
\author{Jiacheng Xie}
\author{Yiyu Zhou}
\author{Yubo Wang}
\author{Chengxing He}
\author{Yu Guo}
\affiliation{Department of Electrical Engineering, Yale University, New Haven, CT 06511, USA}

\author{Hong X. Tang}
\affiliation{Department of Electrical Engineering, Yale University, New Haven, CT 06511, USA}
\affiliation{Corresponding author: hong.tang@yale.edu}
\date{\today}

\begin{abstract}
\vspace{12pt}

Optical parametric oscillators (OPOs) are widely utilized in photonics as classical and quantum light sources. 
Conventional OPOs produce co-propagating signal and idler waves that can be either degenerately or non-degenerately phase-matched. This configuration, however, makes it challenging to separate signal and idler waves and also renders their frequencies highly sensitive to external disturbances. Here, we demonstrate a degeneracy-locked OPO achieved through backward phase matching in a submicron periodically-poled thin-film lithium niobate microresonator. While the backward phase matching establishes frequency degeneracy of the signal and idler, the backscattering in the waveguide further ensures phase-locking between them. Their interplay permits the locking of the OPO's degeneracy over a broad parameter space, resulting in deterministic degenerate OPO initiation and robust operation against both pump detuning and temperature fluctuations. This work thus provides a new approach for synchronized operations in nonlinear photonics and extends the functionality of optical parametric oscillators. With its potential for large-scale integration, it provides a chip-based platform for advanced applications, such as squeezed light generation, coherent optical computing, and investigations of complex nonlinear phenomena.

\end{abstract}
\maketitle

 \noindent\textbf{Introduction} \\[4mm]

Optical parametric oscillators (OPOs) have long been used as narrow-linewidth, coherent light sources with flexible wavelength tunability, making them indispensable in both classical and quantum applications, such as molecular spectroscopy \cite{CW_pulse_OPO,spec_5_12um,spec_infrared}, quantum information processing \cite{review_squeezed,amir_squeezed,QKD,squeeze_NC} and coherent photonic computing \cite{ISM_PRA,Ising_DOPO,Ising_network, Alireza2014OPOIsing,CDOPO_kerr_gaeta}. The miniaturization of OPOs has been a significant focus of recent research, transitioning from bulky free-space systems \cite{OPO_WGR_bulk,CWOPO_8um_bulk,Kerr_OPO_TJK_bulk,octave_kerrOPO_bulk} to compact, chip-based platforms. Advances in nanofabrication, particularly in $\chi^{(2)}$ materials (\textit{e.g.,} aluminum nitride (AlN) and lithium niobate (LN)) and $\chi^{(3)}$ material (\textit{e.g.}, silicon nitride), have resulted in breakthroughs in reduced threshold power \cite{Juanjuan_OPO,Alex_OPO,OPO_amir,PhCOPO_2021}, enhanced conversion efficiency \cite{Kerr_OPO_kartik}, and expanded wavelength coverage and tunability \cite{OPO_octave_alireza,kartik_greenOPO}. Beyond these performance enhancements, the interplay of OPOs, especially $\chi^{(2)}$-OPO with other nonlinear processes has also enabled rich dynamics such as Pockels soliton microcomb \cite{pokels_comb}, frequency-modulated comb generation\cite{FM_OPO}, and non-equilibrium spectral phase transition \cite{NE_phase_transition}. 

Despite these innovations, nearly all integrated $\chi^{(2)}$-based OPOs rely on co-propagating configurations where signal and idler waves are generated in the same direction due to relative ease of the forward quasi-phase matching engineering \cite{Juanjuan_OPO,OPO_amir,OPO_octave_alireza}. This makes their separation particularly challenging, especially in on-chip implementations. More critically, co-propagating OPOs are highly sensitive to perturbations such as temperature fluctuations or pump frequency drifting, which destabilize signal and idler frequencies. This challenge is particularly exacerbated in degenerate OPOs, due to the strict constraints imposed by energy conservation, phase matching, and simultaneous resonance of the pump and half-harmonic modes \cite{coherence_OPO_1990,tuning_OPO_1991}. Therefore, these systems typically require external feedback for stable operation. On the other hand, robust degenerate OPOs are highly desirable for applications in quantum information processing \cite{review_squeezed,amir_squeezed,QKD,squeeze_NC} and photonic Ising machines \cite{ISM_PRA,Ising_DOPO,Ising_network, Alireza2014OPOIsing,CDOPO_kerr_gaeta}, where stability is crucial. 

Here, we demonstrate a degeneracy-locked optical parametric oscillator, achieved through a counter-propagating scheme enabled by advances in submicron periodically poled thin-film lithium niobate \cite{fengyan_SSHG}. Unlike co-propagating OPOs, the counter-propagating configuration is enabled by backward phase matching and naturally separates signal and idler by their opposing propagation directions. While counter-propagating OPOs have previously been demonstrated in bulk periodically poled Potassium Titanyl Phosphate (KTP) crystals, they lack cavity enhancement and instead rely on distributed parametric gain \cite{BOPO_1966,BOPO_theory,canalias2007mirrorless,COPO_photonpair,mutter2024backwardopo}, resulting in kilowatt level threshold power and limited scalability. By incorporating cavity-enhancement, we achieved counter-propagating OPOs with not only a drastic reduction in threshold power to the milliwatt level but also, for the first time, the degeneracy-locking.

This degeneracy-locking mechanism, enforced by simultaneously backward phase matching and backscattering-induced mode hybridization, inherently locks the signal and idler frequencies to half pump frequency and achieves their mutual phase-locking. This ensures deterministic and robust degenerate OPO operation across a pump detuning range up to 3.3\,GHz. Unlike injection-locking, which typically requires an external seed \cite{parametric_seeding_Delhaye, PMsoliton2015, IL_2015_temperal_tweezing}, or synchronization between parallel nonlinear processes \cite{yang2017counter,zhaoyun_OFD,moille2023kerr,gaeta2018sync_cpring}, our degeneracy-locked oscillator achieves frequency synchronization within a single nonlinear process, eliminating the need for external stabilization or complex implementation. Furthermore, the system exhibits a phase transition between symmetric and asymmetric degenerate OPO states, offering a platform to explore rich nonlinear dynamics. This capability, together with the ultra-low threshold power and robust degeneracy-locking, establishes our degeneracy-locked OPO as a promising platform for applications in quantum optics, photonic Ising machines, and optical sensing.

\noindent\textbf{Results.}

\begin{figure*}[t]
\centering
\includegraphics[trim={0cm 0cm 0cm 0cm},clip,width=1\linewidth]{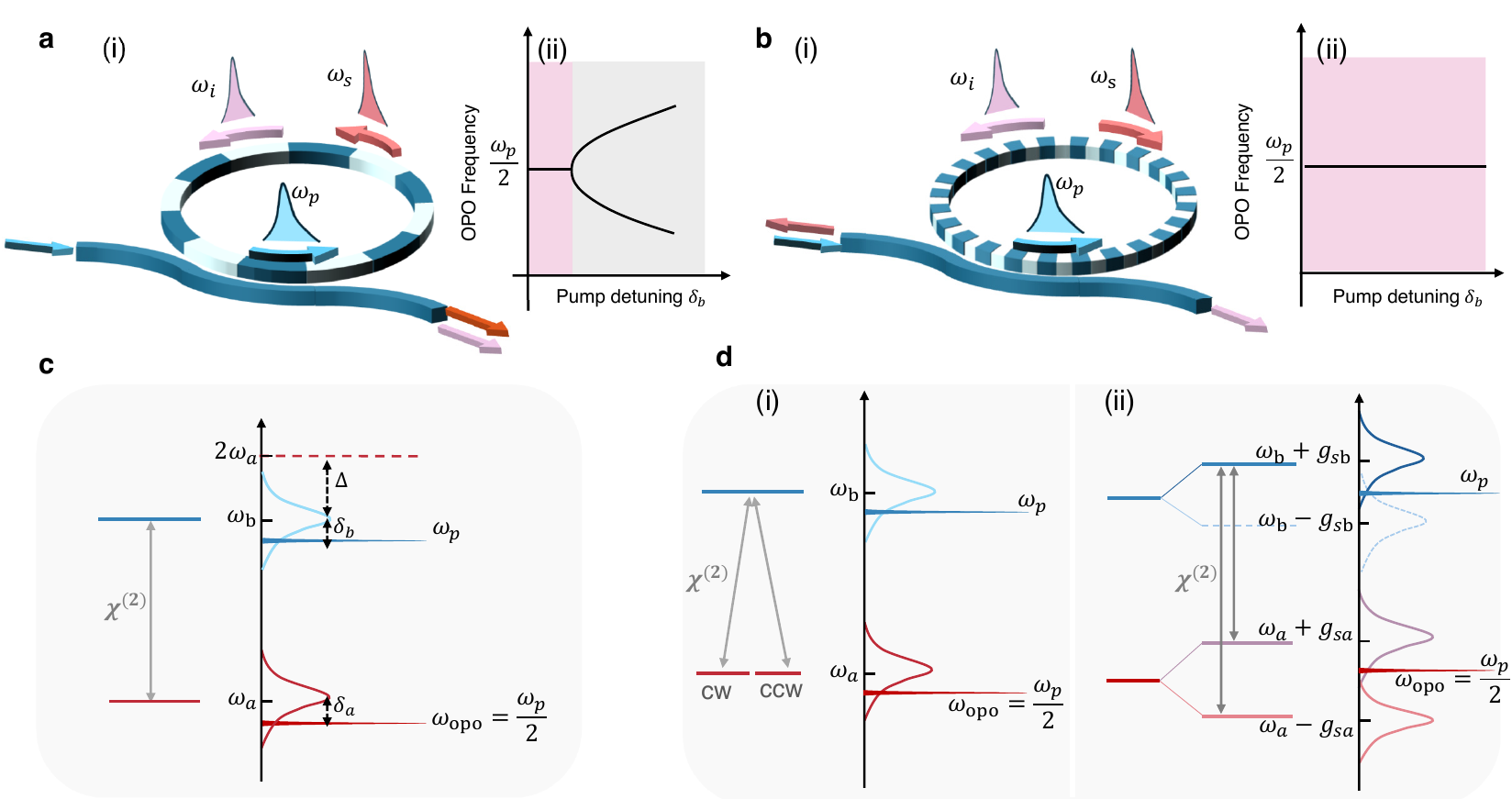}
\caption{{\fontseries{bx}\selectfont Comparison between co-propagating and counter-propagating optical parametric oscillator.} {\fontseries{bx}\selectfont a-b}, (i) Schematic representation of co-propagating (counter-propagating) OPO in micro-ring resonator with longer (shorter) poling period. (ii) The OPO lasing frequency variation against pump detuning is plotted for each scheme. The pink region marks degenerate OPO while the gray marks nondegenerate OPO in (a-ii).
{\fontseries{bx}\selectfont c}, Energy level diagram for the co-propagating degenerate OPO, showing the relationships between pump ($\omega_p$), signal/idler ($\omega_p/2$) frequencies, and their corresponding resonator modes $\omega_b$ and $\omega_a$. The initial resonance mismatch, $\Delta=2\omega_a-\omega_b$, along with the pump detuning $\delta_b=\omega_b-\omega_p$, dictates the signal/idler detuning through $\delta_a=\omega_a-\omega_p/2=(\Delta+\delta_b)/2$.
{\fontseries{bx}\selectfont d}, (i) Energy level diagram for the counter-propagating degenerate OPO where the signal and idler modes are frequency-degenerate but propagate in opposite directions. (ii) Energy level diagram with backscattering, showing the nonlinear coupling between the pump symmetric hybridized mode $b_1$ and the two half-harmonic hybridized modes $a_1$, $a_2$.}
\label{fig1}
\end{figure*}

The conventional co-propagating $\chi^{(2)}$ optical parametric oscillator generates both signal and idler waves in the same direction as the pump wave, as depicted in Fig.\,\ref{fig1}a-(i). The system can be modeled by two interacting photonic energy levels, $\omega_a$ and $\omega_b$, corresponding to two resonator modes, which are coupled via the $\chi^{(2)}$ nonlinearity, as depicted in Fig.\,\ref{fig1}c. This results in the Hamiltonian for the co-propagating degenerate OPO being expressed as $H/\hbar=\delta_a a^\dagger a+\delta_b b^\dagger b+ g_2 a^2b^\dagger+ g_2 (a^\dagger)^2b$, where $a$, $b$ are annihilation operators for the half-harmonic and pump modes, $g_2$ is the second-order nonlinear coupling rate. Here the Hamiltonian is defined in rotating frame $H_{rot}/\hbar=\omega_p b^\dagger b +\frac{\omega_p}{2} a^\dagger a$,  where $\omega_p$ is the pump laser frequency, so $\delta_a$ and $\delta_b$ are corresponding detunings from the mode frequencies. The co-propagating configuration requires both frequency matching and forward momentum matching between the pump, signal and idler waves, expressed as $\omega_p=\omega_s+\omega_i$ and $m_p=m_s+m_i+m_{QPM}$. Here $m_p$, $m_s$ and $m_i$ are the azimuthal modal index for the pump, signal and idler, and $m_{QPM}$ is the quasi-phase matching momentum defined by poling period $\Lambda$ as $m_{QPM}=2\pi R/\Lambda$, with R being the ring radius. 

In co-propagating OPOs, for a given pump mode number $m_p$, if a mode pair $(\{m_s,\omega_s$\},\{$m_i,\omega_i\})$ is both momentum-matched $m_s+m_i+m_{QPM}=m_p$ and frequency-matched $\omega_s+\omega_i=\omega_p$, its neighboring mode pairs $(\{m_s+\mu,\omega_s+\mu\cdot\omega_{FSR}\}$,$\{m_i-\mu,\omega_i-\mu\cdot\omega_{FSR}\})$, $\mu=1,2,3......$ also satisfy both conditions, leading to competition between mode pairs for OPO lasing. The specific mode pair that dominates and leads to OPO emission depends on resonator dispersion and pump detuning \cite{dmitry_Eckhaus,Juanjuan_OPO,OPO_amir}. As a result, the degenerate OPO state typically exists within a narrow pump detuning range, and minute disturbance can drive it to a non-degenerate state, as shown in Fig.\,\ref{fig1}a-(ii). 

In contrast, counter-propagating OPO (CT-OPO) takes a fundamentally different approach, where the signal wave propagates in the opposite direction relative to the idler and pump waves, as shown in Fig.\,\ref{fig1}b-(i). This configuration requires a backward momentum-matching condition $m_p-m_s+m_i=m_{QPM}$, which is challenging to implement due to the need for submicron poling periods \cite{fengyan_SSHG,fengyanScAlN}. In this case, for a given pump mode $m_p$, there is only a single mode pair $(\{m_s,\omega_s$\},\{$m_i,\omega_i\})$ that satisfies both frequency and momentum matching conditions, rendering the CT-OPO inherently free of mode competition.
Specifically, when $m_{QPM}$ is engineered to match $m_p$, the momentum conservation dictates $m_s=m_i$. As a result, the generated signal and idler are degenerate in frequency but propagate in opposite directions—clockwise (cw) for the signal and counterclockwise (ccw) for the idler. The resulting energy levels of the pump, signal and idler modes, as well as their mutual couplings, are depicted in Fig.\,\ref{fig1}d-(i). While the counter-propagating phase-matching ensures frequency degeneracy, the signal and idler reside in separate cavity modes and thus remain phase-unlocked. As a result, they do not exhibit the characteristic random 0 to $\pi$ phase flip relative to the pump, preventing authentic degenerate OPO operation \textcolor{blue}{(see Supplement)}.

The introduction of backscattering-induced linear coupling between the signal mode ($a_{cw}$) and idler mode ($a_{ccw}$) overcomes this limitation and leads to deterministic degenerate OPO generation. The linear coupling between signal and idler modes results in resonance doublets, corresponding to symmetric ($a_1=\frac{a_{cw}+a_{ccw}}{\sqrt{2}}$) and asymmetric ($a_2=\frac{a_{cw}-a_{ccw}}{\sqrt{2}}$) hybridized modes. The splitting between these modes is $2g_{sa}$, where $g_{sa}$ represents the linear coupling rate between the original signal mode $a_{cw}$ and idler mode $a_{ccw}$. Similarly, the pump modes are hybridized into $b_1$ and $b_2$, but phase-matching constraints dictate that only the pump mode $b_1$ with a standing-wave pattern aligned in-phase with the periodically poled domains can achieve nonzero $\chi^{(2)}$ coupling. Therefore, the inclusion of linear coupling leads to the formation of two distinct degenerate OPO states, one occurring between the mode $b_1$ and $a_1$, and another between $b_1$ and $a_2$. The experimentally measured signal ($a_{cw}$) and idler ($a_{ccw}$) emerge exclusively from OPO lasing in either $a_1$ or $a_2$, ensuring both frequency locking at half the pump frequency and mutual phase locking. This phase locking leads to a random binary 0 or $\pi$ phase relationship with respect to the pump \cite{coherence_OPO_1990}, enabling deterministic and robust degenerate OPO operation, as depicted in Fig.\,\ref{fig1}b-(ii).

\begin{figure*}[t]
\centering
\includegraphics[trim={0cm 0cm 0cm 0cm},clip,width=1\linewidth]{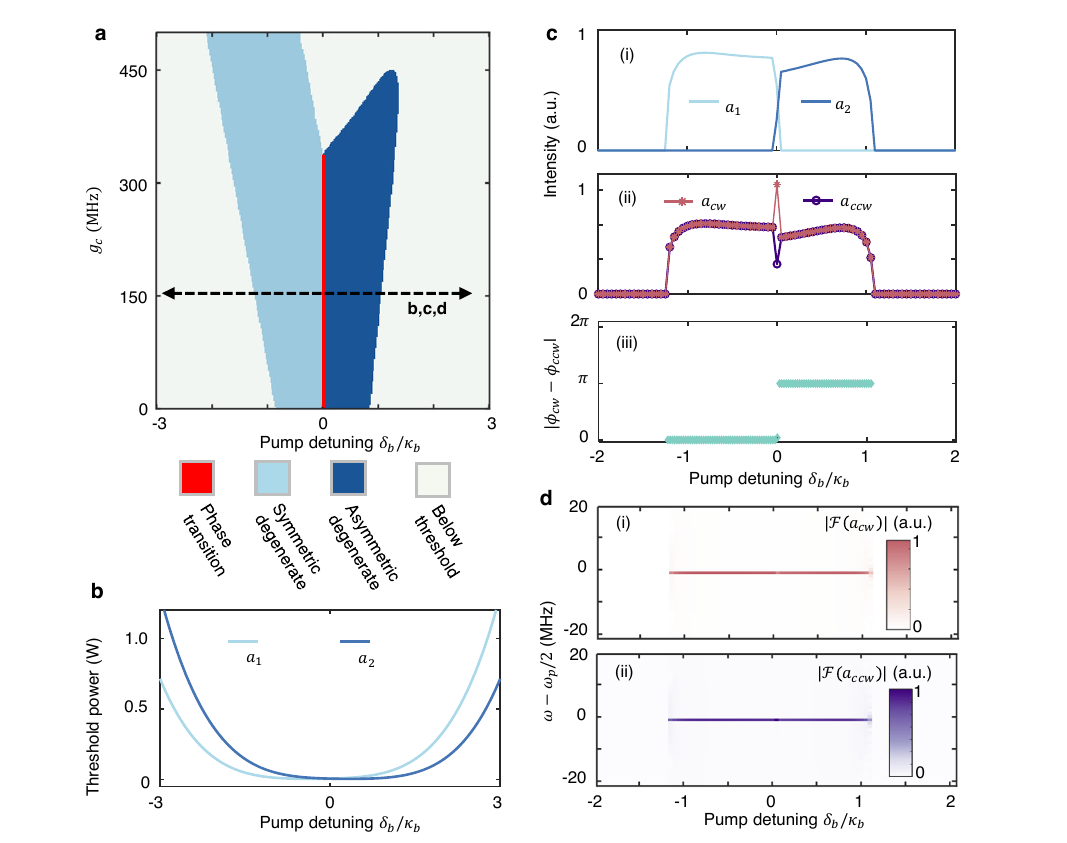}
\caption{{\fontseries{bx}\selectfont Simulated phase diagram and representative behaviors of the degeneracy-locked OPO system. }
{\fontseries{bx}\selectfont a}, The phase space constructed by backscattering coupling rate $g_c$ and pump detuning $\delta_b$, with perfect initial resonance match $\Delta=0$ and a fixed pump power of 5\,mW. Different OPO states are delineated by distinct colors. 
{\fontseries{bx}\selectfont b}, Threshold pump power of symmetric and asymmetric degenerate OPO when $g_c=150$\,MHz. 
{\fontseries{bx}\selectfont c}, For $g_c=150$\,MHz, (i) Intensities for the symmetric hybridized mode $a_{1}$ (light blue) and asymmetric hybridized mode $a_{2}$ (deep blue). (ii) Intensities for the signal mode ($a_{cw}$), plotted by red starred curve and that for idler mode ($a_{ccw}$) plotted by purple circled curve. (iii) Phase difference between $a_{cw}$ and $a_{ccw}$.
{\fontseries{bx}\selectfont d}, Spectrum of $a_{cw}$ and $a_{ccw}$ when $g_c=150$\,MHz at varied pump detunings. 
}
\label{fig2}
\end{figure*}

To understand the dynamics of the system, we calculate the phase diagram of the OPO states by solving the coupling mode equations derived from the system Hamiltonian. \textcolor{blue}{(see supplement)}. The nonlinear coupling rate $g_2$ is calibrated experimentally using SHG measurement to be 0.2\,MHz and adopted here for simulation. \textcolor{blue}{(see supplement)}. The calculations assume perfect initial resonance match ($\Delta=2\omega_a-\omega_b=0$), equal linear coupling rate at both half-harmonic and pump wavelengths ($g_c\equiv g_{sa}=g_{sb}$), and explore the interplay between the linear coupling rate $g_{c}$ and pump detuning $\delta_b$. The resulting phase diagram is presented in Fig.\,\ref{fig2}a.

In the hybridized mode picture, the symmetric ($a_{1}=(a_{cw}+a_{ccw})/\sqrt{2}$) and asymmetric ($ a_{2}=(a_{cw}-a_{ccw})/\sqrt{2}$) hybridized modes undergo $\chi^{(2)}$ interaction with the pump mode $b_1$ for OPO lasing, each possessing distinct thresholds \textcolor{blue}{(see supplement)}. These two thresholds can be adjusted by varying the pump detuning, as depicted by Fig.\,\ref{fig2}b. The OPO state with the lower threshold dominates, clamping the intracavity pump power and preventing the other from initiating. Consequently, the system operates as either a symmetric degenerate OPO in the hybridized mode $a_1$ (light blue area in Fig.\,\ref{fig2}a) or asymmetric degenerate OPO in hybridized mode $a_2$ (dark blue area in Fig.\,\ref{fig2}a). 

To further illustrate the system's behavior, Fig.\,\ref{fig2}c-(i) presents a specific case where $g_c=150$\,MHz (outlined by black dotted lines in Fig.\,\ref{fig2}a). The intracavity photon numbers are observed to populate exclusively in either $a_1$ or $a_2$, corresponding to lasing solely in one of the hybridized modes. Fig.\,\ref{fig2}c-(ii) shows the resulted experimentally measurable outputs, where the signal ($a_{cw}$) and idler ($a_{ccw}$) originate purely from the dominant hybridized OPO and exhibit identical photon numbers except at the phase transition point. Notably, the signal and idler frequencies remain degenerate at $\omega_p/2$, independent of pump detuning $\delta_b$, as shown by the Fourier transformed spectrum of $a_{cw,ccw}$ in Fig.\,\ref{fig2}d. 

As the pump detuning varies, the system undergoes a phase transition from symmetric to asymmetric degenerate OPO states in the regime where linear coupling rate is relatively small. At the transition point where the thresholds of both hybridized OPOs are equal, the intensities for signal and idler undergo splitting. The magnitude of this splitting is enhanced by increased linear coupling, while stronger nonlinear interaction and higher pump power reduce it (\textcolor{blue}{see supplement}). Additionally, the relative phase between the output signal and idler shifts abruptly from $0$ to $\pi$ at the transition point, as depicted in Fig.\,\ref{fig2}c-(iii). This phase transition underscores the dynamic reconfigurability of the hybridized OPO states, further demonstrating the robustness of degenerate OPO operation in this system. As linear coupling increases, pump mode $b_1$ shifts in the opposite direction further from the asymmetric  mode $a_2$ and leads to larger resonance misalignment, and thus raising the OPO threshold and favoring only the symmetric $b_1-a_1$ OPO state (Fig.\,\ref{fig2}a). Nevertheless, increasing the pump power broadens the permissible range of linear coupling, thereby enabling both OPO states to coexist at higher coupling rates.(\textcolor{blue}{see supplement})

\begin{figure*}[t]
\centering
\includegraphics[trim={0cm 0cm 0cm 0cm},clip,width=\linewidth]{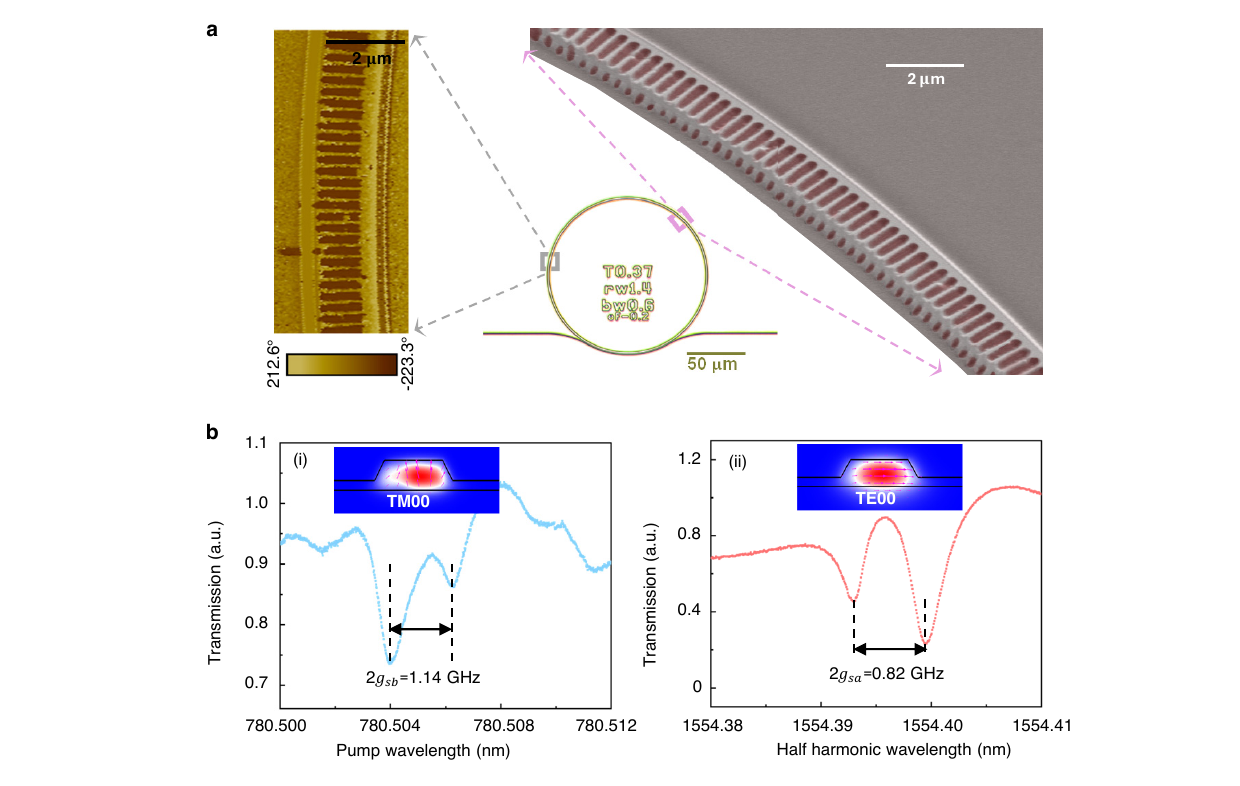}
\caption{{\fontseries{bx}\selectfont Device design and characterization.} 
%{\fontseries{bx}\selectfont a}, Experiment set-up. Red photon detectors: infrared wavelength, Blue: visible wavelength, SiA: silicon absorber. 
{\fontseries{bx}\selectfont a},  Micro-ring resonator (central), Piezoresponse Force Microscope (PFM) image for the ring, showing the phase contrast between the poled and unpoled region (left), and false color Scanning Electron Microscope (SEM) image of the ring after removing the reversed domain by hydrofluoric acid (right). The poling period is 371\,nm.
{\fontseries{bx}\selectfont b}, Transmission spectrum at the pump (i) and half-harmonics (ii) wavelengths, showing corresponding mode splitting. The optical mode profiles for TM00 pump mode and TE00 half-harmonics mode are presented as insets.}
\label{fig_device}
\end{figure*}

To demonstrate the degeneracy-locked OPO and its simulated dynamics experimentally, we implement the CT-OPO in thin-film lithium niobate micro-ring resonators with a precisely engineered periodic poling period of $\Lambda=$371\,nm. This design is specifically tailored for degenerate operation at a pump wavelength around 775 nm as $\Lambda=\frac{\lambda_p}{n_{eff,p}}$, enabling the conversion of the pump TM00 mode to the half-harmonic TE00 mode via the $d_{31}$ nonlinear coefficient. Details of the micro-ring resonator fabrication are provided in Methods. In Fig.\,\ref{fig_device}a, the left panel shows a zoomed-in piezoresponse force microscope (PFM) phase measurement, mapping out the periodically inverted domains with $\Lambda=$371\,nm, while the right panel displays a scanning electron microscope (SEM) image illustrating the volumetric domain structure after the inverted domains are etched by hydrofluoric acid. We first calibrate the backscattering coupling rate by measuring the mode splitting. During the fabrication process, domains with opposite polarization are etched at slightly different rates, creating surface morphology variations that enhance light scattering. The backscattering coupling rates for pump and half-harmonic modes are thus measured to be 570\,MHz and 410\,MHz, as shown in Fig.\,\ref{fig_device}b. 

\begin{figure*}[t]
\centering
\includegraphics[trim={0cm 0cm 0cm 0cm},clip,width=1\linewidth]{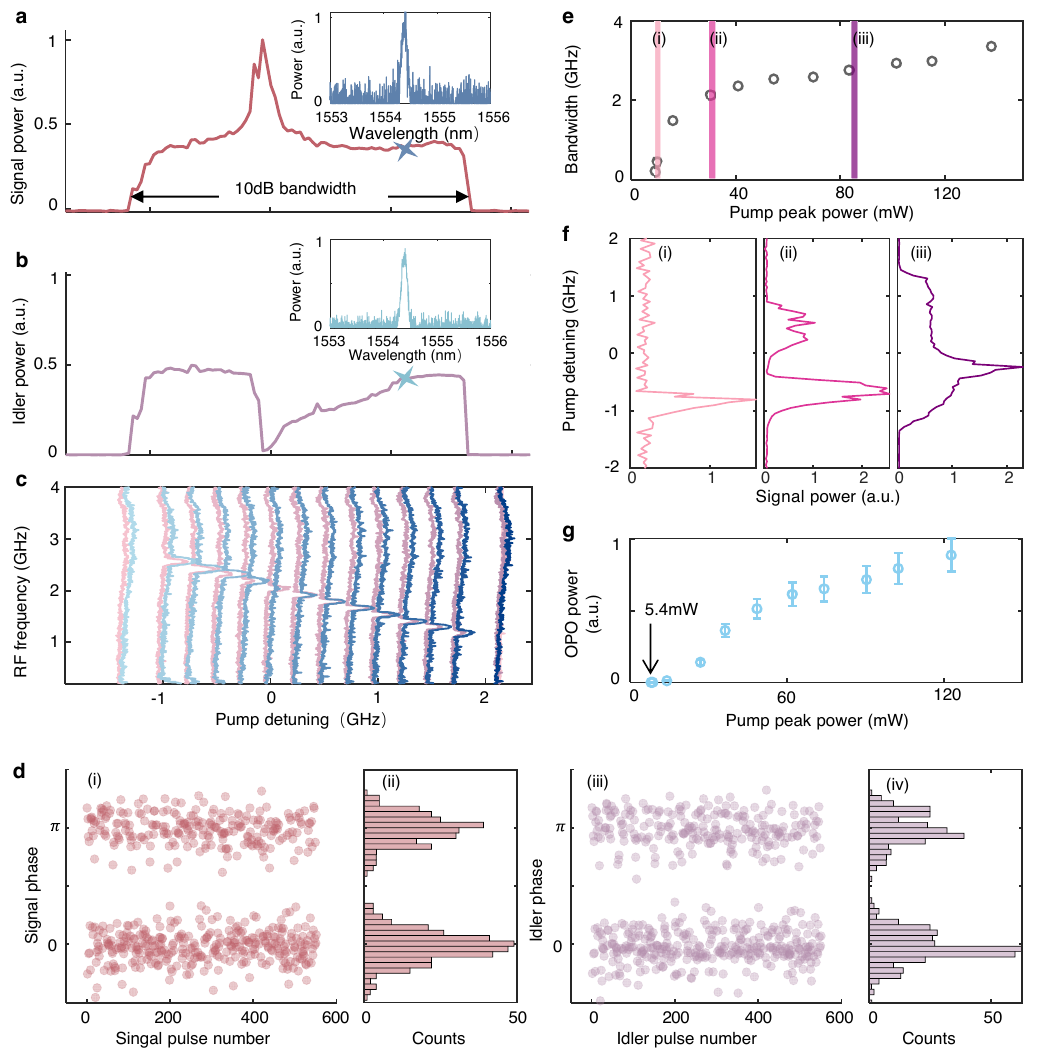}
\caption{{\fontseries{bx}\selectfont Measurement results of synchronized optical parametric oscillator.} 
%{\fontseries{bx}\selectfont a}, Experiment set-up. Red photon detectors: infrared wavelength, Blue: visible wavelength, SiA: silicon absorber. 
{\fontseries{bx}\selectfont a-b}, Optical output power of signal and idler as pump detuning changes.The insets shows corresponding OSA data for emission at 1554.34\,nm. 
{\fontseries{bx}\selectfont c}, Beating note between signal/idler and the local oscillator with fixed frequency as pump detuning changes.  
{\fontseries{bx}\selectfont d}, Relative phase between consecutive signal  pulses (i) and idler pulses (iii), and their statistics (ii,iv). 
{\fontseries{bx}\selectfont e}, The 10dB bandwidth defined through the power spectrum of signal. The visible pump detuning range that enables CT-OPO is twice of the 10dB bandwidth and can be as large as 3.3\,GHz.
{\fontseries{bx}\selectfont f}, The signal power spectrum evolution as pump power changes.
{\fontseries{bx}\selectfont g}, The sum of on-chip signal and idler peak power as on-chip pump peak power. Error bars are deviations considering a coupling fluctuation of 2$\%$ and 5$\%$ for the respective signal/idler and pump light.
}
\label{fig4}
\end{figure*}

In order to efficiently ignite OPO, initial resonance mismatch $\Delta$ between pump and half-harmonic modes is first optimized via temperature tuning to get the maximum SHG outputs \textcolor{blue}{(see supplement)}. The OPO is then driven by an amplified laser pulse with a duration of 15\,ns and a repetition rate of 5\,kHz \textcolor{blue}{(see supplement)}. The optical power spectra of the signal and idler at a pump peak power of $\sim$100\,mW and scanning speed of 1\,nm/s are simultaneously recorded using two identical infrared photon detectors positioned on opposite sides of the device \textcolor{blue}{(see supplement)}. The simultaneously recorded optical power spectra in Fig.\,\ref{fig4}a-b with comparable power levels
demonstrates that signal and idler are indeed counter-propagating. Insets in Fig.\,\ref{fig4}a-b display the corresponding optical spectrum analyzer (OSA) data for a specific pump detuning. Meanwhile, the phase transition from symmetric to asymmetric hybridized OPO is evident from the power splitting of the spectra, where the peak in the signal spectrum (Fig.\,\ref{fig4}a) corresponds to a dip in the idler spectrum (Fig.\,\ref{fig4}b), consistent with the simulation in Fig.\,\ref{fig2}c-(ii). This behavior indicates a first-order phase transition, which typically exhibits sharper features than the softer second-order changes typically observed in isolated co-propagating OPOs \cite{Spectral_PT_2021}. In our case, the theory predicts an abrupt 0-$\pi$ shift in the signal–idler relative phase at the transition point (Fig.\,\ref{fig2}c-(iii)), highlighting the stark discontinuity characteristic of first-order transitions. Recent investigations into non-equilibrium spectral phase transitions in coupled optical parametric oscillators emphasize that such first-order behavior can induce sudden spectral reorganizations and heightened responsiveness to perturbations, resulting in enhanced sensitivity for advanced optical sensing \cite{NE_phase_transition}. 

The frequency degeneracy of the CT-OPO is confirmed via radio-frequency(RF) measurements. When the generated signal and idler are mixed with the same local oscillator at a fixed frequency, the resulting beat notes for the signal (pinkish) and idler (bluish) overlap precisely across various pump detunings, as shown in Fig.\,\ref{fig4}c. The robust degenerate operation of the CT-OPO is ensured by its broad pump detuning range, which renders the system resilient to frequency drifts of the cavity mode and effectively prevents transitions to non-degenerate states or other dynamical regimes such as breathers \cite{OPO_amir, dmitry_Eckhaus}. The degeneracy range is defined by the 10\,dB bandwidth of the signal spectrum and has been experimentally measured to extend up to 3.3\,GHz. The signal spectra at varying pump powers, shown in Fig.\,\ref{fig4}f, reveal a clear broadening of the degeneracy bandwidth with increasing pump power. The ultimate pump detuning range available for degeneracy operation in experiment is limited by the thermal and photon-refractive effect under high power, which shifts resonance away and elevates threshold power thus supressing the OPO emission. 

The random 0-$\pi$ phase flip of the degenerate OPO is further validated by measuring the phase differences between consecutive signal pulses (or idler pulses). To perform this measurement, a 1550\,nm seed laser is frequency-shifted by 300 MHz using an acousto-optic modulator (AOM), then split into two arms and mixed with the signal and idler pulses to extract their phases \textcolor{blue}{(see supplement)}. Fig.\,\ref{fig4}d displays the phase difference results for 600 pulses, collected in sets of 20 pulses and repeated 30 times over a 30-minute period. The phase difference exhibits a clustered distribution around 0 or $\pi$, confirming the continuous operation of the degenerate OPO without any active stabilization feedback applied to the system over a long time.

\begin{figure*}[t]
\centering
\includegraphics[trim={0cm 0cm 0cm 0cm},clip,width=1\linewidth]{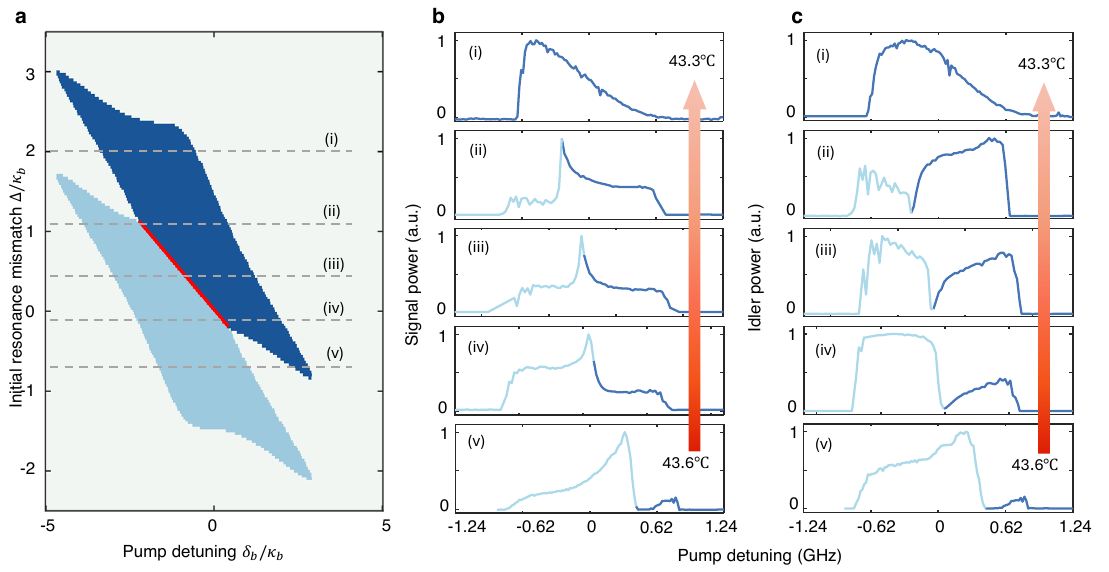}
\caption{{\fontseries{bx}\selectfont Influence of initial resonance mismatch} 
{\fontseries{bx}\selectfont a}, The simulated phase diagram constructed by initial resonance mismatch $\Delta$ and pump detuning $\delta_b$ at a fixed pump power of 15\,mW and backscattering coupling rate $g_{sa}=$410\,MHz, $g_{sb}=$570\,MHz. Colorcode is the same as Fig.\,\ref{fig2}.
{\fontseries{bx}\selectfont b-c}, The signal/idler spectrum evolution as temperature increase, light blue line represents the symmetric degenerate OPO and dark blue line represents the asymmetric one.
}
\label{fig5}
\end{figure*}

Furthermore, the threshold pump power for CT-OPO is significantly reduced—by six orders of magnitude comparing to waveguide-based CT-OPO \cite{BOPO_waveguide} — owing to resonator enhancement and it is measured to be as low as 5.4\,mW, as shown in Fig.\,\ref{fig4}g. This value aligns well with theoretical predictions based on the extracted nonlinear coupling rate $g_2$ from symmetric SHG measurement \textcolor{blue}{(see supplement)}. Due to stringent backward phase matching and backscattering induced synchronization, the CT-OPO consistently locks into a degenerate state above the threshold. The combination of low threshold pump power, deterministic initiation of degenerate OPO, and robust  operation over a broad pump dynamic range establishes counter-propagating OPO as an ideal platform for photonic Ising machines.

Finally, the tuning of the initial resonance mismatch parameter $\Delta$ plays a crucial role in optimizing CT-OPO performance and exploring its rich nonlinear dynamics. By varying $\Delta$ through temperature adjustments, the resonance alignment between the pump and half-harmonic modes can be controlled, influencing the relative dominance of symmetric and asymmetric hybridized OPO states. This capability is demonstrated in the simulated phase diagram (Fig.\,\ref{fig5}a), where the proportion of symmetric and asymmetric degenerate OPO regions shifts as $\Delta$ is tuned. Experimentally, the impact of $\Delta$ on the power spectra of the signal and idler is shown in Fig.\,\ref{fig5}b-c. As $\Delta$ is adjusted via temperature, the phase transition point shifts, and the relative power distributions of symmetric and asymmetric degenerate OPO states follow the trends predicted in the simulation. Despite shifts in dominance between symmetric and asymmetric OPO states, the system still remains in degenerate state, demonstrating its robustness against temperature variations. The ability to tune the initial resonance mismatch $\Delta$ provides a pathway to explore phase-dependent nonlinear dynamics and holds promise for applications in quantum simulation and nonlinear optical sensing.

In addition, when the pump power is below the threshold, our device can also work as a spontaneous parametric down-conversion source to generate spectrally indistinguishable photon pairs. While additional efforts are needed to separate the degenerate photons from a conventional co-propagating SPDC source \cite{he2015ultracompact,2014_onchip_photon_pair_interference,QuanSplit_HOM2007}, the natural spatial separation of the signal and idler in CT-OPO makes our device a potential platform for on-chip quantum optics experiments.

In conclusion, we have demonstrated a degeneracy-locked counter-propagating optical parametric oscillator on a chip-scale platform using submicron periodically poled thin-film lithium niobate. The degeneracy operation is enabled by the interplay of stringent backward quasi-phase matching, which ensures the selection of a single signal-idler mode pair, and backscattering-induced mode hybridization, which locks the system into one of the two degenerate OPO states. This mechanism stabilizes the operation across a broad pump detuning range and ensures resilience to external perturbations such as temperature variations. The hybridized mode picture captures the system dynamics, including phase transitions between symmetric and asymmetric degenerate states, revealing the system’s reconfigurable nature. With its low threshold power, deterministic initiation and robust  operation, and scalable design, the degeneracy-locked OPO provides a robust platform for integrated photonic applications. This work introduces a new mechanism to degenerate OPOs in nonlinear photonics, offering a scalable and energy-efficient platform for advancing next-generation photonic technologies.

\noindent\textbf{Methods} \\
\noindent\textbf{Device fabrication} \\
The LN microresonator is fabricated on a commercial 600\,nm-thick z-cut lithium niobate thin film obtained from NanoLN. The starting material consists of a 50\,nm-thick NbN layer on top of a silicon substrate, followed by a 2\,$\mu$m PECVD SiO$_2$ layer. A 600\,nm z-cut LN layer is then bonded to this substrate. Poling electrodes are patterned using e-beam lithography with a bi-layer PMMA-MMA resist, followed by nickel deposition and a liftoff process. Voltage up to 650\,V lasting for 10\,ms is applied to the electrodes for poling.  The nickel electrodes are subsequently removed using hydrochloric acid. The ring resonators are fabricated following the lithium niobate recipe previously reported in \cite{fengyan_SSHG}. Post-fabrication annealing is conducted at 200$^\circ$C for 12 hours.

\vspace{4 mm}
\noindent\textbf{Data availability} 
The data that support the findings of this study are available from the corresponding authors upon reasonable request.

\vspace{4 mm}
\noindent\textbf{Code availability} 
All relevant computer codes supporting this study are available from the corresponding author upon reasonable request.

\vspace{4 mm}
\noindent \textbf{Acknowledgements.} This work is supported by DARPA through its INSPIRED program under cooperative agreement D24AC00180. The part of the research that involves lithium niobate thin film preparation is supported by the US Department of Energy Co-design Center for Quantum Advantage (C2QA) under Contract No. DE-SC0012704. We thank our cleanroom staff Yong Sun, Lauren McCabe, Yeongjae Shin, Kelly Woods and Michael Rooks for assistance with device fabrication. 

\vspace{4 mm}
\noindent \textbf{Author contributions.} H.X.T. and  F.Y. conceived the experiment. F.Y. fabricated the device. F.Y. performed the experiment. F.Y. analyzed the data. J.X., Y.Z., Y.W., C.H. and Y.G. helped with the project. F.Y., J.X., Y.Z. wrote the manuscript, and all authors contributed to the manuscript. H.X.T. supervised the work.

\vspace{4 mm}
\noindent \textbf{Competing interests.} The authors declare no competing interests.
%\vspace{2 mm}
%\bibliographystyle{myaipnum4-1}
\bibliographystyle{naturemag}
%\vspace{1cm}
\newpage
\def\bibsection{\section{\textbf{references}}}
\bibliography{Ref2}

\begin{thebibliography}{10}
\expandafter\ifx\csname url\endcsname\relax
  \def\url#1{\texttt{#1}}\fi
\expandafter\ifx\csname urlprefix\endcsname\relax\def\urlprefix{URL }\fi
\providecommand{\bibinfo}[2]{#2}
\providecommand{\eprint}[2][]{\url{#2}}

\bibitem{CW_pulse_OPO}
\bibinfo{author}{Dunn, M.~H.} \& \bibinfo{author}{Ebrahimzadeh, M.}
\newblock \bibinfo{title}{Parametric generation of tunable light from continuous-wave to femtosecond pulses}.
\newblock \emph{\bibinfo{journal}{Science}} \textbf{\bibinfo{volume}{286}}, \bibinfo{pages}{1513--1517} (\bibinfo{year}{1999}).

\bibitem{spec_5_12um}
\bibinfo{author}{Maidment, L.}, \bibinfo{author}{Schunemann, P.~G.} \& \bibinfo{author}{Reid, D.~T.}
\newblock \bibinfo{title}{Molecular fingerprint-region spectroscopy from 5 to 12 mum using an orientation-patterned gallium phosphide optical parametric oscillator}.
\newblock \emph{\bibinfo{journal}{Opt. Lett.}} \textbf{\bibinfo{volume}{41}}, \bibinfo{pages}{4261--4} (\bibinfo{year}{2016}).

\bibitem{spec_infrared}
\bibinfo{author}{Arslanov, D.~D.} \emph{et~al.}
\newblock \bibinfo{title}{Continuous‐wave optical parametric oscillator based infrared spectroscopy for sensitive molecular gas sensing}.
\newblock \emph{\bibinfo{journal}{Laser Photonics Rev.}} \textbf{\bibinfo{volume}{7}}, \bibinfo{pages}{188--206} (\bibinfo{year}{2012}).

\bibitem{review_squeezed}
\bibinfo{author}{Andersen, U.~L.}, \bibinfo{author}{Gehring, T.}, \bibinfo{author}{Marquardt, C.} \& \bibinfo{author}{Leuchs, G.}
\newblock \bibinfo{title}{30 years of squeezed light generation}.
\newblock \emph{\bibinfo{journal}{Phys. Scr.}} \textbf{\bibinfo{volume}{91}} (\bibinfo{year}{2016}).

\bibitem{amir_squeezed}
\bibinfo{author}{Amir}.
\newblock \bibinfo{title}{Single-mode squeezed-light generation and tomography with an integrated optical parametric oscillator}.
\newblock \emph{\bibinfo{journal}{Sci. Adv.}}  (\bibinfo{year}{2024}).

\bibitem{QKD}
\bibinfo{author}{Madsen, L.~S.}, \bibinfo{author}{Usenko, V.~C.}, \bibinfo{author}{Lassen, M.}, \bibinfo{author}{Filip, R.} \& \bibinfo{author}{Andersen, U.~L.}
\newblock \bibinfo{title}{Continuous variable quantum key distribution with modulated entangled states}.
\newblock \emph{\bibinfo{journal}{Nat. Commun.}} \textbf{\bibinfo{volume}{3}}, \bibinfo{pages}{1083} (\bibinfo{year}{2012}).

\bibitem{squeeze_NC}
\bibinfo{author}{Zhang, Y.} \emph{et~al.}
\newblock \bibinfo{title}{Squeezed light from a nanophotonic molecule}.
\newblock \emph{\bibinfo{journal}{Nat. Commun.}} \textbf{\bibinfo{volume}{12}}, \bibinfo{pages}{2233} (\bibinfo{year}{2021}).

\bibitem{ISM_PRA}
\bibinfo{author}{Wang, Z.}, \bibinfo{author}{Marandi, A.}, \bibinfo{author}{Wen, K.}, \bibinfo{author}{Byer, R.~L.} \& \bibinfo{author}{Yamamoto, Y.}
\newblock \bibinfo{title}{Coherent ising machine based on degenerate optical parametric oscillators}.
\newblock \emph{\bibinfo{journal}{Phys. Rev. A}} \textbf{\bibinfo{volume}{88}}, \bibinfo{pages}{063853} (\bibinfo{year}{2013}).

\bibitem{Ising_DOPO}
\bibinfo{author}{Inagaki, T.} \emph{et~al.}
\newblock \bibinfo{title}{Large-scale ising spin network based on degenerate optical parametric oscillators}.
\newblock \emph{\bibinfo{journal}{Nat. Photonics}} \textbf{\bibinfo{volume}{10}}, \bibinfo{pages}{415--419} (\bibinfo{year}{2016}).

\bibitem{Ising_network}
\bibinfo{author}{Yamamoto, Y.} \emph{et~al.}
\newblock \bibinfo{title}{Coherent ising machines—optical neural networks operating at the quantum limit}.
\newblock \emph{\bibinfo{journal}{Npj Quantum Inf.}} \textbf{\bibinfo{volume}{3}} (\bibinfo{year}{2017}).

\bibitem{Alireza2014OPOIsing}
\bibinfo{author}{Marandi, A.}, \bibinfo{author}{Wang, Z.}, \bibinfo{author}{Takata, K.}, \bibinfo{author}{Byer, R.~L.} \& \bibinfo{author}{Yamamoto, Y.}
\newblock \bibinfo{title}{Network of time-multiplexed optical parametric oscillators as a coherent ising machine}.
\newblock \emph{\bibinfo{journal}{Nat. Photonics}} \textbf{\bibinfo{volume}{8}}, \bibinfo{pages}{937--942} (\bibinfo{year}{2014}).

\bibitem{CDOPO_kerr_gaeta}
\bibinfo{author}{Okawachi, Y.} \emph{et~al.}
\newblock \bibinfo{title}{Demonstration of chip-based coupled degenerate optical parametric oscillators for realizing a nanophotonic spin-glass}.
\newblock \emph{\bibinfo{journal}{Nat. Commun.}} \textbf{\bibinfo{volume}{11}}, \bibinfo{pages}{4119} (\bibinfo{year}{2020}).

\bibitem{OPO_WGR_bulk}
\bibinfo{author}{Furst, J.~U.} \emph{et~al.}
\newblock \bibinfo{title}{Low-threshold optical parametric oscillations in a whispering gallery mode resonator}.
\newblock \emph{\bibinfo{journal}{Phys. Rev. Lett.}} \textbf{\bibinfo{volume}{105}}, \bibinfo{pages}{263904} (\bibinfo{year}{2010}).

\bibitem{CWOPO_8um_bulk}
\bibinfo{author}{Meisenheimer, S.-K.}, \bibinfo{author}{Fürst, J.~U.}, \bibinfo{author}{Buse, K.} \& \bibinfo{author}{Breunig, I.}
\newblock \bibinfo{title}{Continuous-wave optical parametric oscillation tunable up to an 8$\mu$m wavelength}.
\newblock \emph{\bibinfo{journal}{Optica}} \textbf{\bibinfo{volume}{4}} (\bibinfo{year}{2017}).

\bibitem{Kerr_OPO_TJK_bulk}
\bibinfo{author}{Kippenberg, T.~J.}, \bibinfo{author}{Spillane, S.~M.} \& \bibinfo{author}{Vahala, K.~J.}
\newblock \bibinfo{title}{Kerr-nonlinearity optical parametric oscillation in an ultrahigh-q toroid microcavity}.
\newblock \emph{\bibinfo{journal}{Phys. Rev. Lett.}} \textbf{\bibinfo{volume}{93}}, \bibinfo{pages}{083904} (\bibinfo{year}{2004}).

\bibitem{octave_kerrOPO_bulk}
\bibinfo{author}{Sayson, N. L.~B.} \emph{et~al.}
\newblock \bibinfo{title}{Octave-spanning tunable parametric oscillation in crystalline kerr microresonators}.
\newblock \emph{\bibinfo{journal}{Nat. Photonics}} \textbf{\bibinfo{volume}{13}}, \bibinfo{pages}{701--706} (\bibinfo{year}{2019}).

\bibitem{Juanjuan_OPO}
\bibinfo{author}{Lu, J.} \emph{et~al.}
\newblock \bibinfo{title}{Ultralow-threshold thin-film lithium niobate optical parametric oscillator}.
\newblock \emph{\bibinfo{journal}{Optica}} \textbf{\bibinfo{volume}{8}} (\bibinfo{year}{2021}).

\bibitem{Alex_OPO}
\bibinfo{author}{Bruch, A.~W.}, \bibinfo{author}{Liu, X.}, \bibinfo{author}{Surya, J.~B.}, \bibinfo{author}{Zou, C.-L.} \& \bibinfo{author}{Tang, H.~X.}
\newblock \bibinfo{title}{On-chip $\chi^{(2)}$ microring optical parametric oscillator}.
\newblock \emph{\bibinfo{journal}{Optica}} \textbf{\bibinfo{volume}{6}}, \bibinfo{pages}{1361} (\bibinfo{year}{2019}).

\bibitem{OPO_amir}
\bibinfo{author}{McKenna, T.~P.} \emph{et~al.}
\newblock \bibinfo{title}{Ultra-low-power second-order nonlinear optics on a chip}.
\newblock \emph{\bibinfo{journal}{Nat. Commun.}} \textbf{\bibinfo{volume}{13}}, \bibinfo{pages}{4532} (\bibinfo{year}{2022}).

\bibitem{PhCOPO_2021}
\bibinfo{author}{Marty, G.}, \bibinfo{author}{Combri{\'e}, S.}, \bibinfo{author}{Raineri, F.} \& \bibinfo{author}{De~Rossi, A.}
\newblock \bibinfo{title}{Photonic crystal optical parametric oscillator}.
\newblock \emph{\bibinfo{journal}{Nat. Photonics}} \textbf{\bibinfo{volume}{15}}, \bibinfo{pages}{53--58} (\bibinfo{year}{2021}).

\bibitem{Kerr_OPO_kartik}
\bibinfo{author}{Perez, E.~F.} \emph{et~al.}
\newblock \bibinfo{title}{High-performance kerr microresonator optical parametric oscillator on a silicon chip}.
\newblock \emph{\bibinfo{journal}{Nat. Commun.}} \textbf{\bibinfo{volume}{14}} (\bibinfo{year}{2023}).

\bibitem{OPO_octave_alireza}
\bibinfo{author}{Marandi, A.}
\newblock \bibinfo{title}{Octave-spanning tunable infrared parametric oscillators in nanophotonics}.
\newblock \emph{\bibinfo{journal}{Sci. Adv.}}  (\bibinfo{year}{2023}).

\bibitem{kartik_greenOPO}
\bibinfo{author}{Sun, Y.} \emph{et~al.}
\newblock \bibinfo{title}{Advancing on-chip kerr optical parametric oscillation towards coherent applications covering the green gap}.
\newblock \emph{\bibinfo{journal}{Light Sci. Appl.}} \textbf{\bibinfo{volume}{13}}, \bibinfo{pages}{201} (\bibinfo{year}{2024}).

\bibitem{pokels_comb}
\bibinfo{author}{Bruch, A.~W.} \emph{et~al.}
\newblock \bibinfo{title}{Pockels soliton microcomb}.
\newblock \emph{\bibinfo{journal}{Nat. Photonics}} \textbf{\bibinfo{volume}{15}}, \bibinfo{pages}{21--27} (\bibinfo{year}{2020}).

\bibitem{FM_OPO}
\bibinfo{author}{Stokowski, H.~S.} \emph{et~al.}
\newblock \bibinfo{title}{Integrated frequency-modulated optical parametric oscillator}.
\newblock \emph{\bibinfo{journal}{Nature}} \textbf{\bibinfo{volume}{627}}, \bibinfo{pages}{95--100} (\bibinfo{year}{2024}).

\bibitem{NE_phase_transition}
\bibinfo{author}{Roy, A.}, \bibinfo{author}{Nehra, R.}, \bibinfo{author}{Langrock, C.}, \bibinfo{author}{Fejer, M.} \& \bibinfo{author}{Marandi, A.}
\newblock \bibinfo{title}{Non-equilibrium spectral phase transitions in coupled nonlinear optical resonators}.
\newblock \emph{\bibinfo{journal}{Nat. Phys.}} \textbf{\bibinfo{volume}{19}}, \bibinfo{pages}{427--434} (\bibinfo{year}{2023}).

\bibitem{coherence_OPO_1990}
\bibinfo{author}{Nabors, C.~D.}, \bibinfo{author}{Yang, S.~T.}, \bibinfo{author}{Day, T.} \& \bibinfo{author}{Byer, R.~L.}
\newblock \bibinfo{title}{Coherence properties of a doubly resonant monolithic optical parametric oscillator}.
\newblock \emph{\bibinfo{journal}{J. Opt. Soc. Am. B}} \textbf{\bibinfo{volume}{7}}, \bibinfo{pages}{815--820} (\bibinfo{year}{1990}).

\bibitem{tuning_OPO_1991}
\bibinfo{author}{Eckardt, R.~C.}, \bibinfo{author}{Nabors, C.~D.}, \bibinfo{author}{Kozlovsky, W.~J.} \& \bibinfo{author}{Byer, R.~L.}
\newblock \bibinfo{title}{Optical parametric oscillator frequency tuning and control}.
\newblock \emph{\bibinfo{journal}{J. Opt. Soc. Am. B}} \textbf{\bibinfo{volume}{8}}, \bibinfo{pages}{646--667} (\bibinfo{year}{1991}).

\bibitem{fengyan_SSHG}
\bibinfo{author}{Yang, F.}, \bibinfo{author}{Lu, J.}, \bibinfo{author}{Shen, M.}, \bibinfo{author}{Yang, G.} \& \bibinfo{author}{Tang, H.~X.}
\newblock \bibinfo{title}{Symmetric second-harmonic generation in sub-wavelength periodically poled thin film lithium niobate}.
\newblock \emph{\bibinfo{journal}{Optica}} \textbf{\bibinfo{volume}{11}}, \bibinfo{pages}{1050--1055} (\bibinfo{year}{2024}).

\bibitem{BOPO_1966}
\bibinfo{author}{Harris, S.~E.}
\newblock \bibinfo{title}{Proposed backward wave oscillation in the infrared}.
\newblock \emph{\bibinfo{journal}{Appl. Phys. Lett.}} \textbf{\bibinfo{volume}{9}}, \bibinfo{pages}{114--116} (\bibinfo{year}{1966}).

\bibitem{BOPO_theory}
\bibinfo{author}{Godard, A.}, \bibinfo{author}{Guionie, M.}, \bibinfo{author}{Dherbecourt, J.-B.}, \bibinfo{author}{Melkonian, J.-M.} \& \bibinfo{author}{Raybaut, M.}
\newblock \bibinfo{title}{Backward optical parametric oscillator threshold and linewidth studies}.
\newblock \emph{\bibinfo{journal}{J. Opt. Soc. Am. B}} \textbf{\bibinfo{volume}{39}} (\bibinfo{year}{2022}).

\bibitem{canalias2007mirrorless}
\bibinfo{author}{Canalias, C.} \& \bibinfo{author}{Pasiskevicius, V.}
\newblock \bibinfo{title}{Mirrorless optical parametric oscillator}.
\newblock \emph{\bibinfo{journal}{Nat. Photonics}} \textbf{\bibinfo{volume}{1}}, \bibinfo{pages}{459--462} (\bibinfo{year}{2007}).

\bibitem{COPO_photonpair}
\bibinfo{author}{Luo, K.~H.} \emph{et~al.}
\newblock \bibinfo{title}{Counter-propagating photon pair generation in a nonlinear waveguide}.
\newblock \emph{\bibinfo{journal}{Opt. Express}} \textbf{\bibinfo{volume}{28}}, \bibinfo{pages}{3215--3225} (\bibinfo{year}{2020}).

\bibitem{mutter2024backwardopo}
\bibinfo{author}{Mutter, P.}, \bibinfo{author}{Laurell, F.}, \bibinfo{author}{Pasiskevicius, V.} \& \bibinfo{author}{Zukauskas, A.}
\newblock \bibinfo{title}{Backward wave optical parametric oscillation in a waveguide}.
\newblock \emph{\bibinfo{journal}{npj Nanophotonics}} \textbf{\bibinfo{volume}{1}}, \bibinfo{pages}{38} (\bibinfo{year}{2024}).

\bibitem{parametric_seeding_Delhaye}
\bibinfo{author}{Papp, S.~B.}, \bibinfo{author}{Del’Haye, P.} \& \bibinfo{author}{Diddams, S.~A.}
\newblock \bibinfo{title}{Parametric seeding of a microresonator optical frequency comb}.
\newblock \emph{\bibinfo{journal}{Opt. Express}} \textbf{\bibinfo{volume}{21}}, \bibinfo{pages}{17615--17624} (\bibinfo{year}{2013}).

\bibitem{PMsoliton2015}
\bibinfo{author}{Taheri, H.}, \bibinfo{author}{Eftekhar, A.~A.}, \bibinfo{author}{Wiesenfeld, K.} \& \bibinfo{author}{Adibi, A.}
\newblock \bibinfo{title}{Soliton formation in whispering-gallery-mode resonators via input phase modulation}.
\newblock \emph{\bibinfo{journal}{IEEE Photon. J.}} \textbf{\bibinfo{volume}{7}}, \bibinfo{pages}{1--9} (\bibinfo{year}{2015}).

\bibitem{IL_2015_temperal_tweezing}
\bibinfo{author}{Jang, J.~K.}, \bibinfo{author}{Erkintalo, M.}, \bibinfo{author}{Coen, S.} \& \bibinfo{author}{Murdoch, S.~G.}
\newblock \bibinfo{title}{Temporal tweezing of light through the trapping and manipulation of temporal cavity solitons}.
\newblock \emph{\bibinfo{journal}{Nat. Commun.}} \textbf{\bibinfo{volume}{6}}, \bibinfo{pages}{7370} (\bibinfo{year}{2015}).

\bibitem{yang2017counter}
\bibinfo{author}{Yang, Q.-F.}, \bibinfo{author}{Yi, X.}, \bibinfo{author}{Yang, K.~Y.} \& \bibinfo{author}{Vahala, K.}
\newblock \bibinfo{title}{Counter-propagating solitons in microresonators}.
\newblock \emph{\bibinfo{journal}{Nat. Photonics}} \textbf{\bibinfo{volume}{11}}, \bibinfo{pages}{560--564} (\bibinfo{year}{2017}).

\bibitem{zhaoyun_OFD}
\bibinfo{author}{Zhao, Y.} \emph{et~al.}
\newblock \bibinfo{title}{All-optical frequency division on-chip using a single laser}.
\newblock \emph{\bibinfo{journal}{Nature}} \textbf{\bibinfo{volume}{627}}, \bibinfo{pages}{546--552} (\bibinfo{year}{2024}).

\bibitem{moille2023kerr}
\bibinfo{author}{Moille, G.} \emph{et~al.}
\newblock \bibinfo{title}{Kerr-induced synchronization of a cavity soliton to an optical reference}.
\newblock \emph{\bibinfo{journal}{Nature}} \textbf{\bibinfo{volume}{624}}, \bibinfo{pages}{267--274} (\bibinfo{year}{2023}).

\bibitem{gaeta2018sync_cpring}
\bibinfo{author}{Jang, J.~K.} \emph{et~al.}
\newblock \bibinfo{title}{Synchronization of coupled optical microresonators}.
\newblock \emph{\bibinfo{journal}{Nat. Photonics}} \textbf{\bibinfo{volume}{12}}, \bibinfo{pages}{688--693} (\bibinfo{year}{2018}).

\bibitem{dmitry_Eckhaus}
\bibinfo{author}{Puzyrev, D.~N.} \& \bibinfo{author}{Skryabin, D.~V.}
\newblock \bibinfo{title}{Ladder of eckhaus instabilities and parametric conversion in chi(2) microresonators}.
\newblock \emph{\bibinfo{journal}{Commun. Phys.}} \textbf{\bibinfo{volume}{5}} (\bibinfo{year}{2022}).

\bibitem{fengyanScAlN}
\bibinfo{author}{Yang, F.} \emph{et~al.}
\newblock \bibinfo{title}{{Domain control and periodic poling of epitaxial ScAlN}}.
\newblock \emph{\bibinfo{journal}{Appl. Phys. Lett.}} \textbf{\bibinfo{volume}{123}}, \bibinfo{pages}{101103} (\bibinfo{year}{2023}).

\bibitem{Spectral_PT_2021}
\bibinfo{author}{Roy, A.}, \bibinfo{author}{Jahani, S.}, \bibinfo{author}{Langrock, C.}, \bibinfo{author}{Fejer, M.} \& \bibinfo{author}{Marandi, A.}
\newblock \bibinfo{title}{Spectral phase transitions in optical parametric oscillators}.
\newblock \emph{\bibinfo{journal}{Nat. Commun.}} \textbf{\bibinfo{volume}{12}}, \bibinfo{pages}{835} (\bibinfo{year}{2021}).

\bibitem{BOPO_waveguide}
\bibinfo{author}{Mutter, P.}, \bibinfo{author}{Laurell, F.}, \bibinfo{author}{Pasiskevicius, V.} \& \bibinfo{author}{Zukauskas, A.}
\newblock \bibinfo{title}{Backward wave optical parametric oscillation in a waveguide}.
\newblock \emph{\bibinfo{journal}{npj Nanophotonics}} \textbf{\bibinfo{volume}{1}} (\bibinfo{year}{2024}).

\bibitem{he2015ultracompact}
\bibinfo{author}{He, J.} \emph{et~al.}
\newblock \bibinfo{title}{Ultracompact quantum splitter of degenerate photon pairs}.
\newblock \emph{\bibinfo{journal}{Optica}} \textbf{\bibinfo{volume}{2}}, \bibinfo{pages}{779--782} (\bibinfo{year}{2015}).

\bibitem{2014_onchip_photon_pair_interference}
\bibinfo{author}{Silverstone, J.~W.} \emph{et~al.}
\newblock \bibinfo{title}{On-chip quantum interference between silicon photon-pair sources}.
\newblock \emph{\bibinfo{journal}{Nat. Photonics}} \textbf{\bibinfo{volume}{8}}, \bibinfo{pages}{104--108} (\bibinfo{year}{2014}).

\bibitem{QuanSplit_HOM2007}
\bibinfo{author}{Chen, J.}, \bibinfo{author}{Lee, K.~F.} \& \bibinfo{author}{Kumar, P.}
\newblock \bibinfo{title}{Deterministic quantum splitter based on time-reversed hong-ou-mandel interference}.
\newblock \emph{\bibinfo{journal}{Phys. Rev. A}} \textbf{\bibinfo{volume}{76}}, \bibinfo{pages}{031804} (\bibinfo{year}{2007}).

\end{thebibliography}

\end{document}

\title{Supplementary Information}
\maketitle

\author{} %leave this blank
%% DO NOT ADD AUTHOR INFORMATION HERE; IT WILL BE ADDED DURING PRODUCTION
\onecolumngrid

\tableofcontents

\section{Modeling.}
\subsection{Numerical simulation of the counter-propagating OPO}
Defining the rotating frame as
\begin{align}
    H_{rot}/h=\omega_p(b_{cw}^\dagger b_{cw}+b_{ccw}^\dagger b_{ccw})+\frac{\omega_p}{2}(a_{cw}^\dagger a_{cw}+a_{ccw}^\dagger a_{ccw}),
\end{align}
where $a_{cw,ccw}$ are annihilation operators for half-harmonic signal and idler modes, $b_{cw,ccw}$ are those for pump modes, and $\omega_p$ is pump laser frequency. The Hamiltonian of the counter-propagating OPO in this frame can be written as
\begin{align}
   H/\hbar &=\delta_a (a_{cw}^\dagger a_{cw} + a_{ccw}^\dagger a_{ccw})+\delta_b (b_{cw}^\dagger b_{cw}+b_{ccw}^\dagger b_{ccw}) \nonumber\\
   &+ [g_2a_{cw}^\dagger a_{ccw}^\dagger (b_{cw}+b_{ccw})+h.c.] \nonumber\\
   &+  (g_{sa}a_{cw}^\dagger a_{ccw}+g_{sb}b_{cw}^\dagger b_{ccw}+ h.c.) \nonumber\\
   &+ \varepsilon_P (b_{ccw}^\dagger+h.c.),  
   \label{H_ori_method} 
\end{align}
Where pump is injected into the mode $b_{ccw}$. Defining initial resonance mismatch factor as $\Delta=2\omega_a-\omega_b$, where $\omega_a=\omega_{a,cw}=\omega_{a,ccw}$ is the unperturbed cavity resonance for half-harmonic modes and  $\omega_b=\omega_{b,cw}=\omega_{b,ccw}$ is that for pump modes. The detuning for half-harmonic modes can be expressed by $\delta_a=\omega_a-\omega_p/2=\frac{\Delta+\delta_b}{2}$. Here, $g_2$ represents the second-order nonlinear coupling rate, $g_{sa}$ and $g_{sb}$ denote the back scattering coupling rate for half-harmonic and pump modes, and $\varepsilon_P=\sqrt{\frac{2\kappa_{b,ex}P_{in}}{\hbar\omega_b}}$ is pump strength. 

The coupling mode equations are thus calculated via $\frac{da}{dt}=-i[a,H/\hbar]$ to be
\begin{eqnarray}
        \frac{da_{cw}}{dt}&=&-i\delta_a a_{cw}-\kappa_{a}a_{cw}-ig_2 (b_{cw}a_{ccw}^\dagger+b_{ccw}a_{ccw}^\dagger)-ig_{sa}a_{ccw} \nonumber \\
       \frac{da_{ccw}}{dt}&=&-i\delta_a a_{ccw}-\kappa_{a}a_{ccw}-ig_2 (b_{cw}a_{cw}^\dagger+b_{ccw}a_{cw}^\dagger)-ig_{sa}a_{cw} \nonumber\\
        \frac{db_{cw}}{dt}  &=&-i\delta_b b_{cw}-\kappa_{b}b_{cw}-ig_2a_{cw}a_{ccw}-ig_{sb}b_{ccw} \nonumber \\
       \frac{db_{ccw}}{dt}  &=&-i\delta_b b_{ccw}-\kappa_{b}b_{ccw}-ig_2a_{cw}a_{ccw}-ig_{sb}b_{cw}-i\varepsilon_P. 
\end{eqnarray}

The simulation in the main manuscript for Fig.2 and Fig.5 are performed by numerically solving the above CMEs. Simulation parameters for Fig.2 are:  $\Delta=0$, $g_2=0.2$\,MHz, $P_{in}$=5\,mW, $\kappa_b$=641\,MHz, $\kappa_{cw}=\kappa_{ccw}$=192\,MHz, assuming critical coupling for all the modes so $\kappa_{in}=\kappa_{ex}$. 

\subsection{Counter-propagating OPO without backscattering coupling}

\begin{figure}[t]
\renewcommand{\figurename}{Extended Fig.}
\centering
\includegraphics[trim={0cm 0cm 0cm 0cm},clip,width=1\linewidth]{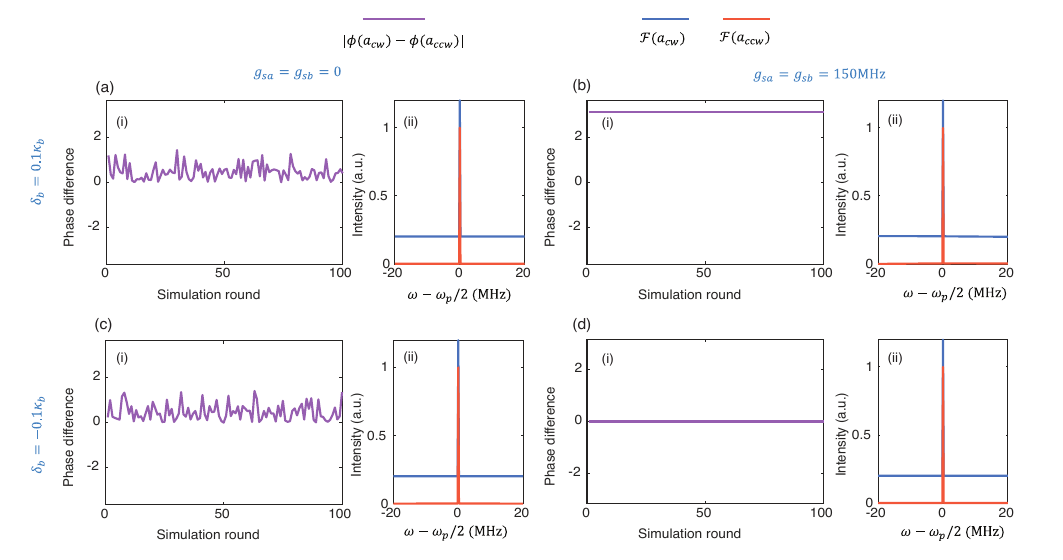}
\caption{
{\fontseries{bx}\selectfont Influence of backscattering on phase locking }
{\fontseries{bx}\selectfont a,} (i) The phase difference betweem signal ($a_{cw}$) and idler ($a_{ccw}$) and (ii) the frequency spectrum for them when linear coupling $g_{sa}=g_{sb}=0$ and pump detuning is $\delta_b=0.1\kappa_b$.
{\fontseries{bx}\selectfont b,} The same plots when $g_{sa}=g_{sb}=150$MHz and $\delta_b=0.1\kappa_b$. 
{\fontseries{bx}\selectfont c,} The same plots when $g_{sa}=g_{sb}=0$ and $\delta_b=-0.1\kappa_b$. 
{\fontseries{bx}\selectfont d,} The same plots when $g_{sa}=g_{sb}=150$MHz and $\delta_b=-0.1\kappa_b$. Other parameters used in simulation: $\Delta=0$, $g_2=$0.2MHz, $P_{in}=5$mW, $\kappa_b$=641MHz, $\kappa_a$=192MHz.
}
\label{EFig1}
\end{figure}

When $g_{sa}=g_{sb}=0$, the OPO is essentially a nondegenerate OPO happening between pump mode $b_{ccw}$ and signal mode $a_{cw}$ and idler mode $a_{ccw}$, the phases between the three lasing signals satisfy $\phi _{s}+\phi_{i}+\pi/2= \phi_p+2n\pi$, where $n=0,1,2...$ . This constraint cannot uniquely determine the value for $\phi_{s}$ and $\phi_{i}$.  If the system is reciprocal so that modes $a_{cw}$ and $a_{ccw}$ share the same physical properties ($\delta_{a,cw}=\delta_{a,ccw}, \kappa_{a,cw}=\kappa_{a,ccw}$), then the generated signal and idler will happen to be frequency-degenerate but still not phase-locked.  In order to demonstrate that counter-propagating phase matching itself gives only frequency degeneracy without phase locking, we solve the above CMEs by setting $g_{sa}=g_{sb}=0$, the simulated phase difference between $a_{cw}$ and $a_{ccw}$ are
presented in Fig.\,\ref{EFig1} (a-i), the variations of $|\phi_{cw}-\phi_{ccw}|$ in different simulation rounds clealy demonstrate their phase-unlocking.  The frequency spectrum shown in  Fig.\,\ref{EFig1} (a-ii) indicates that their frequency are locked to the degeneracy point $\omega_p/2$.  

As a comparison, the phase differences between signal and idler when turning on the linear coupling term show a deterministic value over different simulation rounds, which is either $\pi$ when $\delta_b=0.1\kappa_b$ (Fig.\,\ref{EFig1} b) or $0$ when $\delta_b=-0.1\kappa_b$ (Fig.\,\ref{EFig1} d). 

\subsection{Theoretical analysis of the counter-propagating OPO}

Defining hybridized mode basis as $a_{1}=\frac{a_{cw}+a_{ccw}}{\sqrt{2}},  a_{2}=\frac{a_{cw}-a_{ccw}}{\sqrt{2}}, b_{1}=\frac{b_{cw}+b_{ccw}}{\sqrt{2}}, b_{2}=\frac{b_{cw}-b_{ccw}}{\sqrt{2}}$, the Hamiltonian can be rewritten in the hybridized mode basis as 
\begin{align}
    H_{hybri} /\hbar &= (\delta_a+g_{sa}) a_1^\dagger a_1 +(\delta_a-g_{sa})a_2^\dagger a_2 +(\delta_b+g_{sb})b_1^\dagger b_1 +(\delta_b-g_{sb}) b_2^\dagger b_2 \nonumber \\
     &+\frac{\sqrt{2}g_2}{2}[(a_1^\dagger{}^2-a_2^\dagger{}^2)b_1+h.c.] \nonumber\\
     &+\frac{\sqrt{2}\varepsilon_P}{2}(b_1^\dagger -b_2^\dagger+h.c.).
\end{align}

The nonlinear interaction term in this Hamiltonian reveals two key insights, 1. only the symmetric hybridized pump mode $b_1$ will participate in the $\chi^{(2)}$ interaction, whereas the hybridized pump mode $b_2$ serves as additional pump power dissipation channel. 2. inclusion of linear coupling leads to the formation of two distinct degenerate OPO states, one occurring between the modes $b_1$ and $a_1$, another occurring between $b_1$ and $a_2$. The coupling mode equations (CMEs) are thus calculated via $\frac{da}{dt}=-i[a,H_{hybri}/\hbar]$ to be

\begin{eqnarray}
        \frac{da_{1}}{dt}&=&-i(\delta_a+g_{sa})a_{1}-\kappa_{a1}a_1-i\sqrt{2}g_2 a^\dagger_{1}b_1 \nonumber\\
        \frac{da_{2}}{dt}&=&-i(\delta_a-g_{sa})a_{2}-\kappa_{a2}a_2+
        i\sqrt{2}g_2 a^\dagger_{2}b_1 \nonumber \\
        \frac{db_1}{dt}  &=&-i(\delta_b+g_{sb})b_1-\kappa_{b1}b_1-i\frac{g_2}{\sqrt{2}} (a_{1}^2-a_{2}^2)-i\frac{\varepsilon_P}{\sqrt{2}} \nonumber \\
         \frac{db_2}{dt}  &=&-i(\delta_b-g_{sb})b_2-\kappa_{b2}b_2+i\frac{\varepsilon_P}{\sqrt{2}}.
\end{eqnarray}

% In experiment, the backscattering coupling not only leads to mode splitting but also makes the linewidth of the two hybridized modes different, which means there is dissipative coupling coming into play.
% Here we investigate how the dissipative coupling term influences the phase transition in hybridized OPOs. 
% \\

The above CMEs have two sets of steady-state solution, i.e. $a_2=0, \frac{da_1}{dt}=0,\frac{db_1}{dt}=0$ and $a_1=0, \frac{da_2}{dt}=0,\frac{db_1}{dt}=0$. Analytical analysis for the system can be given based on certain simplifications and assumptions. The thresholds for hybridized OPOs can be analytically solved and compared. Consider the steady OPO emission between $a_1$ and $b_1$ when $a_2=0$, then we have 
\begin{eqnarray}
-i\delta_{a1} a_{1}-\kappa_{a1} a_{1}-i\sqrt{2}g_2 a^\dagger_{1}b_1&=& 0  \\
 -i\delta_{b1} b_1-\kappa_{b1}b_1-i\frac{g_2}{\sqrt{2}}a_{1}^2-i\frac{\varepsilon_P}{\sqrt{2}}&=&0, \label{eq:b_steady}
\end{eqnarray}
where $\delta_{a1}=\delta_a+g_{sa}=\frac{\Delta+\delta_b}{2}+g_{sa}$, $\delta_{b1}=\delta_b+g_{sb}$.
Therefore, when OPO reaches steady-state, $b_1=\frac{-i\delta_{a1}-\kappa_{a1}}{i\sqrt{2}g_2}\frac{a_1}{a_1^\dagger}=\frac{-i\delta_{a1}-\kappa_{a1}}{i\sqrt{2}g_2}e^{i2\phi_1}$, assuming $a_1=|a_1|e^{i\phi_1}$. Substituting this expression for $b_1$ into Eq.\,\ref{eq:b_steady}, we have 
\begin{equation}
    (-i\delta_{b1}-\kappa_{b1})\cdot\frac{-i\delta_{a1}-\kappa_{a1}}{ig_2}-ig_2|a_1|^2=i\varepsilon_Pe^{-i2\phi_1} \label{eq:a1_b}.
\end{equation}
Separating the real and imaginary part of Eq.\,\ref{eq:a1_b} gives
\begin{eqnarray}
    \kappa_{a1}\kappa_{b1}-\delta_{a1}\delta_{b1}+g_2^2|a_1|^2&=&-g_2\varepsilon_P\mathrm{cos}(2\phi_1),\\
    \delta_{a1}\kappa_{b1}+\delta_{b1}\kappa_{a1}&=&g_2\varepsilon_P\mathrm{sin}(2\phi_1).
\end{eqnarray}
Since $|a_1|^2=\frac{\varepsilon_P}{g_2}\cdot\sqrt{1-\frac{(\delta_{a1}\kappa_{b1}+\delta_{b1}\kappa_{a1})^2}{g_2^2\varepsilon_P^2}} - \frac{\kappa_{a1}\kappa_{b1}-\delta_{a1}\delta_{b1}}{g_2^2}+\geq 0$, the resulted $\varepsilon_P$ should satisfy 
\begin{equation}
    \varepsilon_P^2 \geq \frac{(\kappa_{a1}\kappa_{b1}-\delta_{a1}\delta_{b1})^2+(\delta_{a1}\kappa_{b1}+\delta_{b1}\kappa_{a1})^2}{g_2^2}.
\end{equation}

Hence the threshold pump powers for the hybridized OPOs are 
\begin{eqnarray}
       P_{th,i}=\frac{(\kappa_{ai}\kappa_{b1}-\delta_{ai}\delta_{b1})^2+(\delta_{ai}\kappa_{b1}+\delta_{b1}\kappa_{ai})^2}{g_2^2}\cdot \frac{\hbar \omega_p}{2\kappa_{b1,ex}}.
\end{eqnarray}
The subscript $i \in \{1, 2\}$ denotes the indices for the symmetric and asymmetric hybridized half-harmonic modes, respectively. The phase transition occurs when the threshold pump powers are equal, i.e., $P_{th,1} = P_{th,2}$. In the case where the two hybridized modes share the same dissipation rate $\kappa_{a1}=\kappa_{a2}=\kappa_a$ and the liner coupling rates are the same $g_{sa}=g_{sb}=g_c$, the phase transition point is reached when the pump detuning compensates for the initial resonance mismatch, namely,  when $\delta_b = -\Delta$, thus $\delta_{a1,a2}=\pm g_c$, $\delta_{b1}=g_c-\Delta$. At this point, the co-shared threshold power can be simplified into:
\begin{equation}
    P_{th,c}=\frac{\hbar\omega_p}{2\kappa_{ex,b1}}\cdot\frac{[(g_c-\Delta)^2+\kappa_{b1}^2]\cdot(\kappa_a^2+g_{c}^2)}{g_2^2}. 
    \label{eq:threshold_power}
\end{equation}

\begin{figure}[t]
\renewcommand{\figurename}{Extended Fig.}
\centering
\includegraphics[trim={0cm 0cm 0cm 0cm},clip,width=1\linewidth]{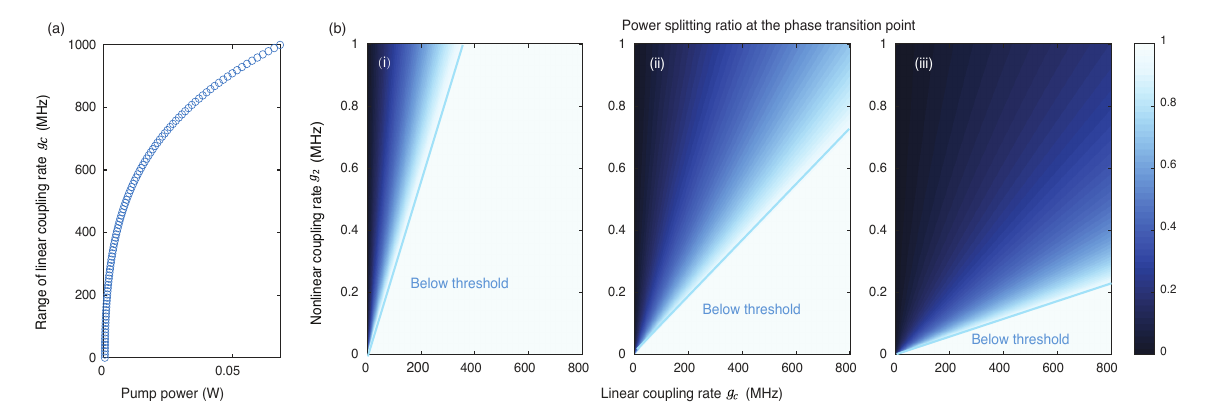}
\caption{ {\fontseries{bx}\selectfont Theoretical calculation results}
{\fontseries{bx}\selectfont a,} (i) The permissible linear coupling range for coexistence of dual degenerate OPO states for different pump power.
{\fontseries{bx}\selectfont b,} The power splitting ratio under different linear and nonlinear coupling rates. Panels (i-iii) depict the situation when applied pump power is 0.1, 1, 10\,mW, respectively.
Other parameters used here are: $\Delta=0$, $g_2=$0.2MHz, $P_{in}=5$mW, $\kappa_{b1}$=641MHz, $\kappa_a$=192MHz.
}
\label{EFig2}
\end{figure}

The occurrence of dual OPO states and associated phase transition point is guaranteed when the applied pump power surpass this co-shared threshold power. A larger linear coupling rate $g_c$ will elevate the required power for dual OPO states. The permissible range of linear coupling for the coexistence of dual OPO states is depicted in Fig.\,\ref{EFig2}(a). 

At the phase transition points, both hybridized OPOs are lasing simultaneously, the resulted output signal $a_{cw}$ and idler $a_{ccw}$ are thus an interference between the lased OPOs. The photon-number splitting ratio $R$ are calculated via 
\begin{equation}
  R=\frac{|a_{cw}|^2-|a_{ccw}|^2 }{2|a_1||a_2|}=\mathrm{cos}(\phi_1-\phi_2),\nonumber 
\end{equation}
where $\phi_i=\frac{1}{2}\mathrm{arcsin}(\frac{\delta_{ai} \kappa_{b1}-\Delta \kappa_{ai}}{g_2\varepsilon_P})$, $\delta_{ai}=(-1)^{i-1}g_{sa}$ $i \in \{1, 2\}$.
The splitting ratio basically increases as linear coupling rate $g_{sa}$ increases while a larger $\chi^{(2)}$ nonlinear coupling rate and a more intense pump lower down the ratio, as illustrated by Fig.\,\ref{EFig2}(b).

\section{Measurement}
\subsection{Experiment setup for OPO measurement}
The pump laser is a custom-built amplified pulsed laser. A continuous wave (CW) seed laser at 1550 nm is shaped into pulses using an intensity electro-optic modulator (EOM) controlled by an arbitrary waveform generator (AWG). The AWG outputs voltages $V_{\text{high}}$ and $V_{\text{low}}$ to switch the EOM between its on and off states. The pulses are then amplified using a two-stage EDFA setup, with a 1 nm bandwidth filter placed between the EDFAs to remove background noise from the first stage and enhance the pulse extinction ratio. The amplified infrared pulses are subsequently frequency-doubled to the visible range using a commercial PPLN crystal [\,Fig.\,\ref{EFig3}(a)\,]. A visible photon detector measures the average power of the pulse $P_{\text{on}}$, and the peak power is calculated using $(P_{\text{on}}-P_{\text{off}})/f_{\text{rep}}W$, where $P_{\text{off}}$ represents the optical power when AWG output is DC off state. In our experiment, the repetition rate $f_{\text{rep}}$ and pulse duration $W$ are set to be 5\,KHz and 15\,ns for the measurements in Fig.4 in the main manuscript.

For OPO generation, the visible pump laser is injected through the 780 nm port of the left wavelength-division multiplexer (WDM), while the generated signal is collected from the 1550\,nm port on the same side. A second WDM on the opposite side separates the transmitted pump light and the generated idler. To prevent residual pump light from interfering with the photodetector (PD) response, silicon absorbers with 40\,dB attenuation at 780\,nm are used. To verify that the signal and idler are generated at degeneracy, a local oscillator, provided by another 1550\,nm laser, is split into two paths and mixed with both the signal and idler for simultaneous beat-note detection. The resulting signals are recorded by two fast (12 GHz) photon detectors and analyzed using an electrical spectrum analyzer (ESA) [\,Fig.\,\ref{EFig3}(b)\,].

\begin{figure}[t]
\renewcommand{\figurename}{Extended Fig.}
\centering
\includegraphics[trim={0cm 0cm 0cm 0cm},clip,width=0.85\linewidth]{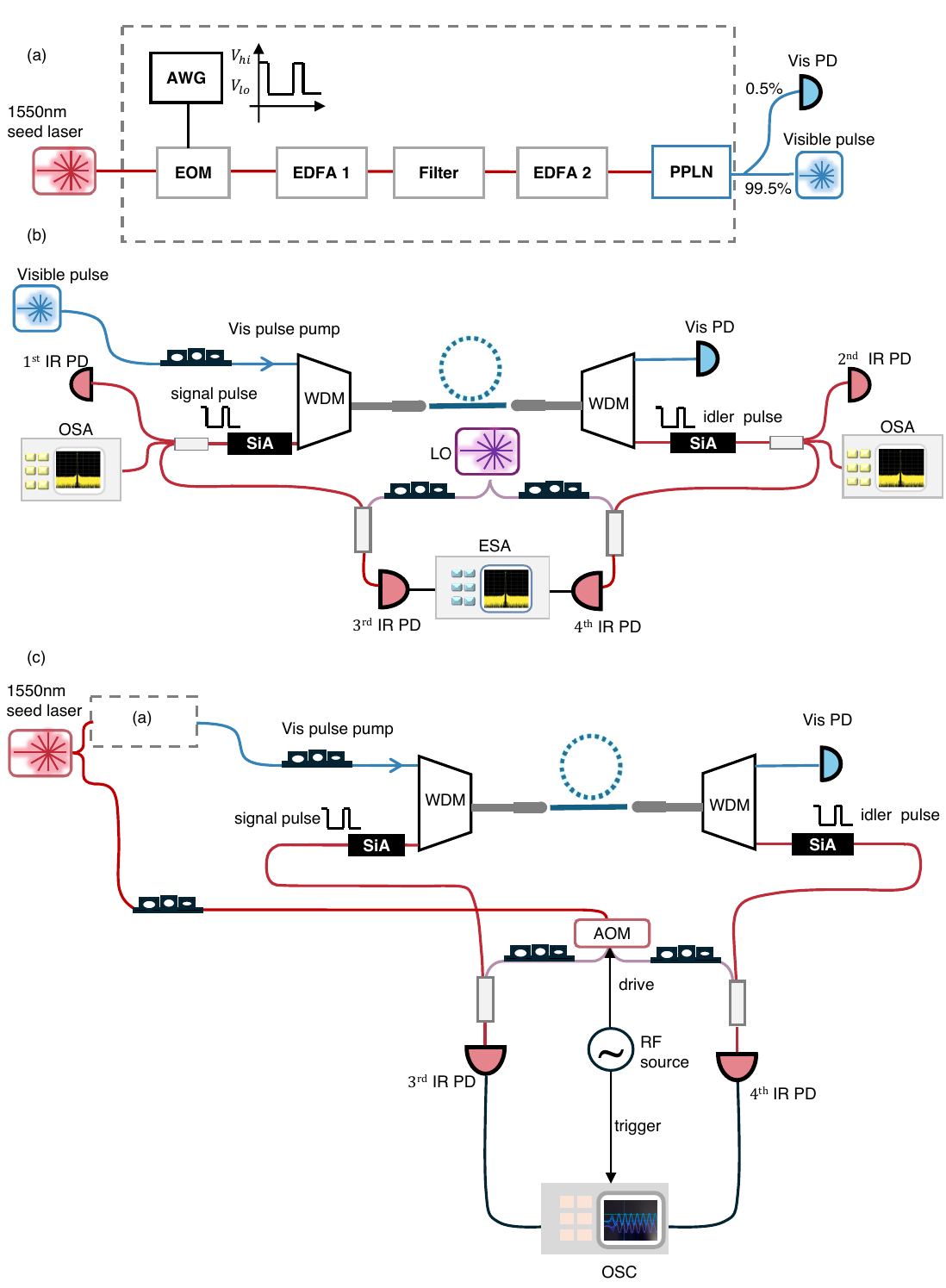}
\caption{
{\fontseries{bx}\selectfont Measurement Setup}
{\fontseries{bx}\selectfont a,} Schematic of the 780\,nm nanosecond pulsed laser system.
{\fontseries{bx}\selectfont b,} Experimental setup for counter-propagating OPO measurement. SiA represents a silicon absorber used to block transmitted or reflected residual 780\,nm pump light. The first and second infrared photon detectors are slow, while the third and fourth detectors are high-speed. The local oscillator is another continuous-wave 1550\,nm laser with a fixed wavelength.
{\fontseries{bx}\selectfont c,} Experimental setup for phase measurement of the degeneracy-locked OPO. Here, the local oscillator is derived from the same 1550 nm seed laser and frequency-shifted by 300 MHz using an AOM.}
\label{EFig3}
\end{figure}

\subsection{Degenerate OPO phase measurement}
To measure the phase differences between consecutive signal pulses in the degenerate OPO, 10$\%$ of the 1550 nm seed laser is extracted and sent to a 300\,MHz acousto-optic modulator (AOM), while the remaining 90$\%$ follows the optical path shown in Fig.\,\ref{EFig3}(a) to generate the 780\,nm pulsed pump.

The AOM-shifted laser is mixed with the generated signal, producing a 300\,MHz beat note with a pulsed envelope. The pump repetition rate is set to 100\,kHz, with a pulse duration of 50\,ns. Since the relative phase between the signal and pump randomly locks to either 0 or $\pi$ upon each OPO initiation, the phase differences between consecutive signal pulses are also expected to be 0 or $\pi$, provided the pump phase remains stable over one pulse period (10\,$\mu$s). An oscilloscope records the beat notes, and Fig.\,\ref{EFig3}(a) displays 20 overlaid consecutive beat notes, showing 300\,MHz oscillations with binary phase relationships.

Phase difference is determined via cross-correlation, with the delay at peak correlation corresponding to the phase changes between successive pulses. Fig.\,\ref{EFig3}(b) presents two example cross-correlation curves, where peak values occur at phase delays of approximately 0 or $\pi$. The distribution centered around 0 and $\pi$ in Fig.\,4(d) of the main manuscript is attributed to pump phase variations during a single pulse period.

The same data acquisition and analysis procedure is applied to idler pulses.

\begin{figure}[t]
\renewcommand{\figurename}{Extended Fig.}
\centering
\includegraphics[trim={0cm 0cm 0cm 0cm},clip,width=1\linewidth]{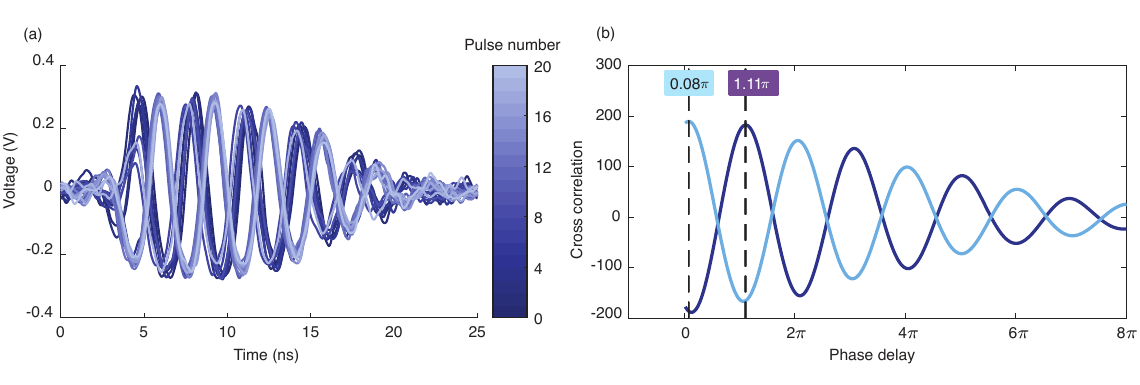}
\caption{
{\fontseries{bx}\selectfont Phase measurement}
{\fontseries{bx}\selectfont a,} Beat notes of 20 consecutive signal pulses mixed with the AOM-shifted laser.
{\fontseries{bx}\selectfont b,}  Cross-correlation of the 4th–5th and 6th–7th beat notes from (a), revealing phase delays of 0.08$\pi$ and 1.11$\pi$, respectively.
}
\label{EFig3}
\end{figure}

\subsection{Symmetric second-harmonic generation}
\begin{figure}[t]
\renewcommand{\figurename}{Extended Fig.}
\centering
\includegraphics[trim={0cm 0cm 0cm 0cm},clip,width=1\linewidth]{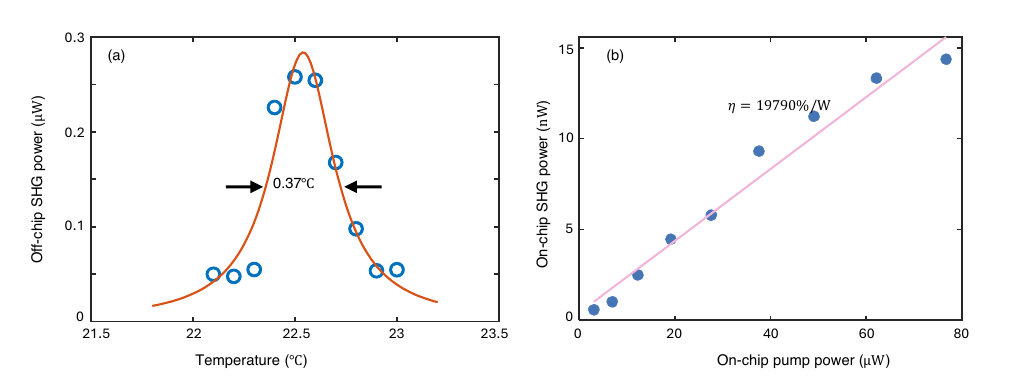}
\caption{
{\fontseries{bx}\selectfont Symmetric SHG measurement}
{\fontseries{bx}\selectfont a,} Symmetric SHG output power as temperature varies. The full-width at half-maximum window is fitted to be 0.37\,$^\circ$
{\fontseries{bx}\selectfont b,} Symmetric SHG conversion efficiency at pump undepleted regime.
}
\label{EFig4}
\end{figure}

To ignite counter-propagating OPO, the initial resonance mismatch $\Delta$ should be optimized, ideally close to zero. Thus, the preliminary resonance alignment between the pump mode ($\omega_b$) and the signal/idler mode ($\omega_a$) is achieved using a threshold-less symmetric second-harmonic generation (SSHG) measurement. In this process, continuous wave 1550\,nm light is injected from both ends of the device, and symmetric SHG outputs are simultaneously monitored. The chip temperature is adjusted using a heater beneath the chip holder, with the optimal operating temperature determined by the peak SHG output. The 3-dB temperature bandwidth for SSHG is approximately 0.37$^\circ$C, as shown in Fig.\,\ref{EFig4}(a). 

Following the initial alignment procedure, we calibrated the nonlinear coupling rate $g_2$ through SSHG power measurements in the pump undepleted regime. By fitting the on-chip left-propagating SSHG power as a function of on-chip pump power, we determined an SHG efficiency of $P_{SHG}/P_{p}^2 = 19790\%/\text{W}$ (Fig.\,\ref{EFig4}(b)), where $P_{SHG}$ and $P_p$ represent the one-sided SHG power and one-sided pump power, respectively. The theoretical expression for the SHG conversion efficiency is given by:
\begin{equation} 
\frac{P_{SHG}}{P_{p}^2} = \frac{g_2^2 \omega_b}{\hbar \omega_p^2} \cdot \frac{2\kappa_{ex,b}}{\delta_b^2 + \kappa_b^2} \cdot \left(\frac{2\kappa_{ex,a}}{\delta_a^2 + \kappa_a^2}\right)^2, \label{eq:SSHG_eff}
\end{equation}
where $g_2$ is the nonlinear coupling rate, $\kappa_{a}$, $\kappa_{ex,a}$, $\kappa_{b}$, and $\kappa_{ex,b}$ are the external and total coupling rates for mode a and b, $\delta_a$ and $\delta_b$ are the mode detunings. By substituting the experimentally fitted values for $\kappa_{ex,a}$, $\kappa_a$, $\kappa_{ex,b}$, and $\kappa_b$ into Eq.\,\ref{eq:SSHG_eff}, and assuming an ideal dual-resonance condition where $\delta_a = \delta_b = 0$, we calculate the coupling rate $g_2$ to be 0.2\,MHz. This result is based on experimentally measured quality factors of $Q_{a,in} = 7.75 \times 10^5$, $Q_{a,ex} = 5.04\times 10^6$, $Q_{b,in} = 6.10\times 10^5$, and $Q_{b,ex} = 8.22\times10^6$. The resulted theoretical threshold power for OPO is thus calculated to be 1.4\,mW according to Eq.\,\ref{eq:threshold_power}, which is consistent in order of magnitude with the experimentally measured value. The nonlinear coupling rate can be further enhanced by utilizing $d_{33}$ for conversion from the 780 nm TM00 mode to the 1560 nm TM00 mode. However, the current device exhibits a lower quality factor for the TM00 mode compared to the TE00 mode, requiring further optimization to improve quality factors for efficient TM00-TM00 conversion.

